\documentclass[a4paper,12pt]{article}

\usepackage{amsmath}
\usepackage{amssymb}
\newlength{\pubnumber} 

\catcode`\@=11
\@addtoreset{equation}{section}

\def\section{\@startsection{section}{1}{\z@}{3.5ex plus 1ex minus .2ex}
 {2.3ex plus .2ex}{\large\bf}}
\def\subsection{\@startsection{subsection}{2}{\z@}{2.3ex plus .2ex}
 {2.3ex plus .2ex}{\bf}}

\begin{document}
\begin{titlepage}
\samepage{
\setcounter{page}{0}
\rightline{OUTP--03--30P}
\rightline{LPTENS-03/36}
\rightline{\tt hep-th/0403058}
\rightline{February 2004}
\vfill
\begin{center}
 {\Large \bf Classification of the chiral \\
        $Z_2\times Z_2$ fermionic models in the
                heterotic superstring\\ }
\vfill
\vspace{.25in}
 {\large A.E. Faraggi$^{1}$, C. Kounnas$^{2}$, S.E.M. Nooij$^{1}$,
   $\,$and$\,$ J. Rizos$^{2,3}$\\}
\vspace{.25in}
{\it $^{1}$ Theoretical Physics Department, University of Oxford,
            Oxford, OX1 3NP, UK\\}
\vspace{.05in}
 {\it  $^{2}$ Laboratoire de Physique Theorique,
Ecole Normale Sup\'erieure, 24 rue L'homond, F--75231 Paris Cedex 05, France \\
}
\vspace{.05in}
 {\it  $^{3}$ Department of Physics,
              University of Ioannina, GR45110 Ioannina, Greece\\}
\end{center}
\vfill
\begin{abstract}
The first particle physics observable whose origin may be sought
in string theory is the triple replication of the matter generations.
The class of $Z_2\times Z_2$ orbifolds of six dimensional compactified
tori, that have been most widely studied in the free fermionic formulation,
correlate the family triplication with the existence of three twisted
sectors in this class.
In this work we seek an improved understanding of the geometrical origin
of the three generation free fermionic models. Using fermionic
and orbifold techniques we classify the $Z_2\times Z_2$ orbifold
with symmetric shifts on six dimensional compactified internal
manifolds. We show that perturbative three generation models
are not obtained in the case of $Z_2\times Z_2$
orbifolds with symmetric shifts on complex tori, and that the
perturbative three generation models in this class necessarily employ an
asymmetric shift. We present a class of three generation models
in which the $SO(10)$ gauge symmetry
cannot be broken perturbatively, while preserving the Standard Model
matter content. We discuss the potential implications of the
asymmetric shift for strong--weak coupling duality and moduli stabilization.
We show that the freedom in the modular invariant
phases in the $N=1$ vacua that
control the chiral content,
can be interpreted as vacuum expectation values
of background fields of the underlying $N=4$ theory, whose
dynamical components are projected out by the $Z_2$--fermionic
projections.
In this class of vacua the chiral content of the models is
determined by the underlying $N=4$ mother theory.

\end{abstract}
\vfill
\smallskip}
\end{titlepage}

\def\beq{\begin{equation}}
\def\eeq{\end{equation}}
\def\beqn{\begin{eqnarray}}
\def\la{\label}
\def\eeqn{\end{eqnarray}}
\def\Tr{{\rm Tr}\,}
\def\KM{{Ka\v{c}-Moody}}

\def\ie{{\it i.e.}}
\def\etc{{\it etc}}
\def\eg{{\it e.g.}}
\def\half{{\textstyle{1\over 2}}}
\def\third{{\textstyle {1\over3}}}
\def\quarter{{\textstyle {1\over4}}}
\def\m{{\tt -}}
\def\p{{\tt +}}

\def\rep#1{{\bf {#1}}}
\def\slash#1{#1\hskip-6pt/\hskip6pt}
\def\slk{\slash{k}}
\def\GeV{\,{\rm GeV}}
\def\TeV{\,{\rm TeV}}
\def\y{\,{\rm y}}
\def\SM{Standard-Model }
\def\SUSY{supersymmetry }
\def\SSM{supersymmetric standard model}
\def\vev#1{\left\langle #1\right\rangle}
\def\l{\langle}
\def\r{\rangle}

\def\Htw{{\tilde H}}
\def\chibar{{\overline{\chi}}}
\def\qbar{{\overline{q}}}
\def\ibar{{\overline{\imath}}}
\def\jbar{{\overline{\jmath}}}
\def\Hbar{{\overline{H}}}
\def\Qbar{{\overline{Q}}}
\def\abar{{\overline{a}}}
\def\alphabar{{\overline{\alpha}}}
\def\betabar{{\overline{\beta}}}
\def\tautwo{{ \tau_2 }}
\def\calF{{\cal F}}
\def\calP{{\cal P}}
\def\calN{{\cal N}}
\def\smallmatrix#1#2#3#4{{ {{#1}~{#2}\choose{#3}~{#4}} }}
\def\bone{{\bf 1}}
\def\V{{\bf V}}
\def\N{{\bf N}}
\def\bQ{{\bf Q}}
\def\t#1#2{{ \Theta\left\lbrack \matrix{ {#1}\cr {#2}\cr }\right\rbrack }}
\def\C#1#2{{ C\left\lbrack \matrix{ {#1}\cr {#2}\cr }\right\rbrack }}
\def\tp#1#2{{ \Theta'\left\lbrack \matrix{ {#1}\cr {#2}\cr }\right\rbrack }}
\def\tpp#1#2{{ \Theta''\left\lbrack \matrix{ {#1}\cr {#2}\cr }\right\rbrack }}
\def\ul#1{$\underline{#1}$}
\def\bE#1{{E^{(#1)}}}
\def\IZ{\relax{\bf Z}}\def\IC{\relax{\bf C}}
\def\IR{\relax{\rm I\kern-.18em R}}
\def\lamb{\lambda}
\def\fc#1#2{{#1\over#2}}
\def\hx#1{{\hat{#1}}}
\def\Gh{\hat{\Gamma}}
\def\subsubsec#1{\noindent {\it #1} \br}
\def\WP{{\bf WP}}
\def\gn{\Gamma_0}
\def\bgn{{\bar \Gamma}_0}
\def\Ds{\Delta^\star}
\def\abstract#1{
\vskip .5in\vfil\centerline
{\bf Abstract}\penalty1000
{{\smallskip\ifx\answ\bigans\leftskip 2pc \rightskip 2pc
\else\leftskip 5pc \rightskip 5pc\fi
\noindent\abstractfont \baselineskip=12pt
{#1} \smallskip}}
\penalty-1000}
\def\us#1{\underline{#1}}
\def\hth/#1#2#3#4#5#6#7{{\tt hep-th/#1#2#3#4#5#6#7}}
\def\nup#1({Nucl.\ Phys.\ $\us {B#1}$\ (}
\def\plt#1({Phys.\ Lett.\ $\us  {B#1}$\ (}
\def\cmp#1({Commun.\ Math.\ Phys.\ $\us  {#1}$\ (}
\def\prp#1({Phys.\ Rep.\ $\us  {#1}$\ (}
\def\prl#1({Phys.\ Rev.\ Lett.\ $\us  {#1}$\ (}
\def\prv#1({Phys.\ Rev.\ $\us  {#1}$\ (}
\def\mpl#1({Mod.\ Phys.\ Let.\ $\us  {A#1}$\ (}
\def\ijmp#1({Int.\ J.\ Mod.\ Phys.\ $\us{A#1}$\ (}
\def\br{\hfill\break}\def\ni{\noindent}
\def\mbr{\hfill\break\vskip 0.2cm}
\def\cx#1{{\cal #1}}\def\al{\alpha}\def\IP{{\bf P}}
\def\ov#1#2{{#1 \over #2}}
\def\be{\beta}\def\al{\alpha}
\def\b{{\bf b}}
\def\S{{\bf S}}
\def\X{{\bf X}}
\def\I{{\bf I}}
\def\mb{{\mathbf b}}
\def\mS{{\mathbf S}}
\def\mX{{\mathbf X}}
\def\mI{{\mathbf I}}
\def\balpha{{\mathbf \alpha}}
\def\bbeta{{\mathbf \beta}}
\def\bgamma{{\mathbf \gamma}}
\def\bxi{{\mathbf \xi}}

\def\ul#1{$\underline{#1}$}
\def\bE#1{{E^{(#1)}}}
\def\IZ{\relax{\bf Z}}\def\IC{\relax{\bf C}}
\def\IR{\relax{\rm I\kern-.18em R}}
\def\lam{\lambda}
\def\fc#1#2{{#1\over#2}}
\def\hx#1{{\hat{#1}}}
\def\Gh{\hat{\Gamma}}
\def\subsubsec#1{\noindent {\it #1} \br}
\def\WP{{\bf WP}}
\def\gn{\Gamma_0}
\def\bgn{{\bar \Gamma}_0}
\def\Ds{\Delta^\star}
\def\abstract#1{
\vskip .5in\vfil\centerline
{\bf Abstract}\penalty1000
{{\smallskip\ifx\answ\bigans\leftskip 2pc \rightskip 2pc
\else\leftskip 5pc \rightskip 5pc\fi
\noindent\abstractfont \baselineskip=12pt
{#1} \smallskip}}
\penalty-1000}
\def\us#1{\underline{#1}}
\def\hth/#1#2#3#4#5#6#7{{\tt hep-th/#1#2#3#4#5#6#7}}
\def\nup#1({Nucl.\ Phys.\ $\us {B#1}$\ (}
\def\plt#1({Phys.\ Lett.\ $\us  {B#1}$\ (}
\def\cmp#1({Commun.\ Math.\ Phys.\ $\us  {#1}$\ (}
\def\prp#1({Phys.\ Rep.\ $\us  {#1}$\ (}
\def\prl#1({Phys.\ Rev.\ Lett.\ $\us  {#1}$\ (}
\def\prv#1({Phys.\ Rev.\ $\us  {#1}$\ (}
\def\mpl#1({Mod.\ Phys.\ Let.\ $\us  {A#1}$\ (}
\def\ijmp#1({Int.\ J.\ Mod.\ Phys.\ $\us{A#1}$\ (}
\def\br{\hfill\break}\def\ni{\noindent}
\def\mbr{\hfill\break\vskip 0.2cm}
\def\cx#1{{\cal #1}}\def\al{\alpha}\def\IP{{\bf P}}
\def\ov#1#2{{#1 \over #2}}
\def\be{\beta}\def\al{\alpha}


\def\inbar{\,\vrule height1.5ex width.4pt depth0pt}

\def\IC{\relax\hbox{$\inbar\kern-.3em{\rm C}$}}
\def\IQ{\relax\hbox{$\inbar\kern-.3em{\rm Q}$}}
\def\IR{\relax{\rm I\kern-.18em R}}
 \font\cmss=cmss10 \font\cmsss=cmss10 at 7pt
\def\IZ{\relax\ifmmode\mathchoice
 {\hbox{\cmss Z\kern-.4em Z}}{\hbox{\cmss Z\kern-.4em Z}}
 {\lower.9pt\hbox{\cmsss Z\kern-.4em Z}}
 {\lower1.2pt\hbox{\cmsss Z\kern-.4em Z}}\else{\cmss Z\kern-.4em Z}\fi}

\def\AEF{A.E. Faraggi}
\def\KRD{K.R. Dienes}
\def\JMR{J. March-Russell}
\def\JHEP#1#2#3{JHEP {\textbf #1}, (#2) #3}
\def\NPB#1#2#3{{\it Nucl.\ Phys.}\/ {\bf B#1} (#2) #3}
\def\PLB#1#2#3{{\it Phys.\ Lett.}\/ {\bf B#1} (#2) #3}
\def\PRD#1#2#3{{\it Phys.\ Rev.}\/ {\bf D#1} (#2) #3}
\def\PRL#1#2#3{{\it Phys.\ Rev.\ Lett.}\/ {\bf #1} (#2) #3}
\def\PRT#1#2#3{{\it Phys.\ Rep.}\/ {\bf#1} (#2) #3}
\def\MODA#1#2#3{{\it Mod.\ Phys.\ Lett.}\/ {\bf A#1} (#2) #3}
\def\IJMP#1#2#3{{\it Int.\ J.\ Mod.\ Phys.}\/ {\bf A#1} (#2) #3}
\def\nuvc#1#2#3{{\it Nuovo Cimento}\/ {\bf #1A} (#2) #3}
\def\etal{{\it et al,\/}\ }

\hyphenation{su-per-sym-met-ric non-su-per-sym-met-ric}
\hyphenation{space-time-super-sym-met-ric}
\hyphenation{mod-u-lar mod-u-lar--in-var-i-ant}

\newcommand{\cc}[2]{c\genfrac{[}{]}{0pt}{}{#1}{#2}}
\newcommand{\oo}[2]{\left(#1\left|#2\right.\right)}
\newcommand{\bartheta}[2]{\bar{\theta}\begin{bmatrix} #1 \\ #2 \end{bmatrix}}
\newcommand{\cption}[1]{Inequivalent realistic liftable models with a #1 gauge
group. The chiral content of each model is listed per plane and
numbered, $+$ lists all the positive chiral states per plane
while $-$ lists all the negative states per plane. The total
sum of all the planes is then listed and subsequently the net
total number of chiral states. The list is ordered by the total
net number of chiral states.}
\def\th{\theta}
\def\bth{\bar{\theta}}
\def\chir{\text{Ch}}

\section{Introduction}\label{one}

String theory is in a precarious state of affairs. On the one hand
the theory shows great promise in its ability to provide
a consistent framework for perturbative quantum gravity,
while at the same time giving rise to the gauge and
matter structures that are observed experimentally.
However, the existence of a multitude of possible string vacua
has led some authors to lose all hope and to advocate
resorting to anthropic principles as the possible resolution
for the contrived set of parameters that seem to govern
our universe \cite{anthropics}.

Our point point of view is different. Ultimately the search
for the principles that underly string theory and the vacuum
selection will entail the conceptual resolution
of the quantum gravity synthesis, and the fundamental
understanding of quantum mechanics with its probablistic
interpretation when applied to the space--time arena.

A more pragmatic view of string theory suggests
that the basic properties
of the low energy data, as well as the basic properties
of string  theories should be utilised in trying to isolate
vacua, or classes of vacua, that look most promising. From the
low energy data point of view we may hypothesise
that the viable string theory vacua should accommodate
two pivotal ingredients: the existence of {\it three generations}
and their {\it embedding in} an underlying $SO(10)$, or $E_6$
grand unified group structure. From the string theory
point of view the basic properties that may serve as guides
are the various T-- and S--duality symmetries.
In this respect it is also plausible that the self--dual
points under these dualities may play a role in the vacuum
selection principle.

A given set of string vacua that exhibits compelling properties
must then be investigated in depth. In the least these can be viewed
as case examples providing the concrete laboratories to
study how the properties of the observed data may arise
from quantum gravity, and to develop the tools to relate
between the theory and experiment. However, there also
exist the possibility that certain case examples capture
some properties of the true string vacuum that may eventually prove
relevant for the understanding of the low energy data.
In any case, it is clear that different approaches must
be pursued for better understanding of string theory
and its possible connection with experimental data.

The first sector among the low energy experimental observables
whose origin we may seek in a theory of quantum
gravity is the flavor sector. In the context of the quantum field
theories underlying the Standard Model of particle physics,
this group of parameters does not arise from any
physical principle, like the gauge principle.
It is then encouraging that already from the early days
of superstring phenomenology, it was observed that
the flavor replication is related a topological
property of the string compactifications, namely the Euler
characteristic \cite{candelas}. However, this observation does not yet
explain the existence of three generations. The first particle
physics observable whose origin we may seek to relate
to string theory is therefore the replication of the three
matter generations.

Among the most advanced string models to date are the three generation
heterotic string models~\cite{fsu5,fny,alr,euslm,lykken,lr},
constructed in the free-fermion
formulation~\cite{fff}.
These models have been the subject of detailed studies, showing
that they can, at least in principle, account for desirable physical
features including the observed fermion mass spectrum, the longevity of
the proton, small neutrino masses, the consistency of gauge-coupling
unification with the experimental data from LEP and elsewhere, and the
universality of the soft supersymmetry-breaking parameters~\cite{reviewsp}.
An important property of the fermionic construction
is the standard $SO(10)$ embedding of the Standard Model spectrum,
which ensures natural consistency with the experimental
values for $\alpha_s(M_Z)$ and $\sin^2\theta_W(M_Z)$.
Furthermore, this class of models
yielded the only known string model that reproduces in the
low energy effective field theory solely the spectrum
of the Minimal Supersymmetric Standard Model \cite{cfn}.

A vital property of the realistic fermionic models
is their underlying $Z_2\times Z_2$
orbifold structure.
Many of the encouraging phenomenological characteristics of these models
are rooted in this structure. In particular, the emergence of the
three chiral generations in a large class of fermionic constructions
is correlated with the existence of three twisted sectors
in the $Z_2\times Z_2$ orbifold of the six dimensional internal
manifold.
Each twisted sector produces exactly one
of the light chiral generations and there is no additional chiral
matter. Thus, the fermionic construction
offers a plausible
and compelling explanation to the existence of three generations
in nature.

To see more precisely the orbifold correspondence of the
fermionic construction,
we recall that the free-fermion models are generated by a set of
basis vectors which define the transformation properties of the
world--sheet fermions as they are transported around
non--contractible loops of the string world sheet. A set of
realistic fermionic models contains a subset of boundary
conditions, the so called extended NAHE--set, which can be seen to
correspond to $Z_2\times Z_2$ orbifold compactification with the
standard embedding of the gauge connection~\cite{foc}.
The fermionic model constructed just with the basis vectors
of the extended NAHE--set gives rise to 24 generations from the
twisted sectors, as well as three additional generation/anti--generation
pairs from  the untwisted sector. At the $N=4$ level the fermionic
point in the moduli space corresponds to an $SO(12)$ enhancement of
the internal lattice. The induced  $Z_2\times Z_2$ action  gives rise to
a model with $(h_{11},h_{21})=(27,3)$, matching the data of
the free-fermion model. We note that the data of this model differs
from the $Z_2\times Z_2$ orbifold at a generic point in the moduli space,
which has $(h_{11},h_{21})=(51,3)$. Alternatively, we can start with the
$Z_2\times Z_2$ orbifold at a generic point and produce the one at
the free fermionic point by adding a freely acting shift on the internal
lattice. \cite{befnq,partitions}.

The above remarks make apparent the need to understand better the
general structure of the realistic free fermionic models,
and, in particular, the geometrical structure that underlies the three
generation models.

In the framework of the fermionic construction
the three generations
are obtained by adding three, or four, additional boundary
condition basis vectors beyond the minimal NAHE--set. The
basis vectors reduce the number of generations to
three generations, one from each of the twisted sectors of the
$Z_2\times Z_2$ orbifold.

In this paper we observe in some
of the concrete quasi--realistic
three generation models \cite{euslm} that the action of two of
the additional boundary condition basis vectors correspond to
symmetric shifts on the internal coordinates, whereas the third corresponds
to a fully asymetric shift.
 We then proceed to classify all possible
$Z_2\times Z_2$ orbifolds with symmetric shifts, and demonstrate
that three generations cannot be obtained solely with symmetric shifts
on complex tori.
This is one of the main results of the analysis and it reveals, at least in
the context of the three generation
models, that the geometrical structures that underly these models
may not be simple Calabi--Yau
manifolds, but it corresponds to geometries that are yet to be defined.
This observation may eventually prove important for the issue of
moduli stabilization.

Additionally, we will demonstrate the existence of three generation models
with a perturbatively unbroken $SO(10)/E_6$ gauge group, in which the
internal manifold is reduced to a product of six circles. This again
demonstrates the possibility that the geometries relating
to the viable vacua may not correspond to the complex geometries
that have been more prevalent in the literature. Some of
phenomenological difficulties that have been associated
with symmetric compactifications, like supersymmetry breaking and
moduli stabilization, may therefore be cured in the viable geometries.
This class of models, while not viable with respect to perturbative
phenomenology, produces one generation from a single fixed point
in each twisted sector. Hence, realizing the $Z_2\times Z_2$
geometric picture of the three chiral generations.
Our classification demonstrates additionally that in a large class
of $N=1$ models the freedom in the phases appearing in the $N=1$
partition function can be understood as the Vacuum Expectation Value (VEV)
of background fields of the $N=4$ underlying theory, whose dynamical
components are projected out by the extra $Z_2\times Z_2$
projections. Thus, the information on the chiral content of the
$N=1$ models is already contained at the $N=4$ level.
Examples of this phenomenon are already noted in the case
of the $Z_2\times Z_2$ orbifold on $SO(12)$ versus $SO(4)^3$
lattices, as discussed above.

Our paper is organised as follows: in section \ref{two}
we discuss the general structure of the models based on the fermionic
construction. In
a concrete model we show that the additional boundary
vectors beyond the NAHE--set can be regarded as two
symmetric shifts plus one fully asymmetric shift.
The main aim of this section is to establish the connection
of the analysis to follow with the phenomenological
three generation models.

In section \ref{three} we present the setup of our analysis.
We present the most
general free fermionic model describing heterotic string with
a $Z_2\times Z_2$ orbifold description.
In section \ref{four} we present our method to classify
all possible symmetric shifts and proceed to perform the complete
classification for gauge groups that descend from the $N=4$ mother theory.
We find that down to six generations the perturbative
models can be described
in terms of symmetric shifts and hence possess a geometrical
interpretation in terms of $Z_2\times Z_2$ symmetric orbifolds.
However, the three generation perturbative models are not admitted
in this classification and entail an additional shift which
is necessarily asymmetric between the left and the right--movers.
We demonstrate the existence of a class of three twisted generation models
in which the GUT symmetry group cannot be broken perturbatively,
while preserving complete twisted matter multiplets.
Additionally, in this class of models the six dimensional
internal lattice is reduced to a product of six circles.
Hence, one of the main conclusions of the the analysis is
that in the framework of $Z_2\times Z_2$ orbifolds,
three generations models are not obtained solely with
symmetric shifts on complex tori, and suggests
that the geometrical objects underlying the quasi--realistic
free fermionic models are more esoteric than ordinary
$Z_2\times Z_2$ Calabi--Yau manifolds.
In section \ref{five} we present our results and
section \ref{conclusion} concludes our paper.

\section{General structure of realistic free fermionic models}\label{two}
In this section we recapitulate the main structure of
the realistic free fermionic models.
The notation
and details of the construction of these
models are given elsewhere \cite{fsu5,fny,alr,euslm,cfn,nahe,cfs}.
In the free fermionic formulation of the heterotic string
in four dimensions all the world-sheet
degrees of freedom  required to cancel
the conformal anomaly are represented in terms of free world--sheet
fermions \cite{fff}.
In the light-cone gauge the world-sheet field content consists
of two transverse left- and right-moving space-time coordinate bosons,
$X^\mu_{1,2}$ and ${\bar X}^\mu_{1,2}$,
and their left-moving fermionic superpartners $\psi^\mu_{1,2}$,
and additional 62 purely internal
Majorana-Weyl fermions, of which 18 are left-moving,
and 44 are right-moving.
In the supersymmetric sector the world-sheet supersymmetry is realised
non-linearly and the world-sheet supercurrent \cite{supercurrent}
is given by
\begin{equation}\label{eq:supercurrent}
T_F=\psi^\mu\partial X_\mu+i\chi^Iy^I\omega^I,~(I=1,\cdots,6).
\end{equation}
The $\{\chi^{I},y^I,\omega^I\}~(I=1,\cdots,6)$ are 18 real free
fermions transforming as the adjoint representation of ${\rm SU}(2)^6$.
Under parallel transport around a non-contractible loop on the toroidal
world-sheet the fermionic fields pick up a phase,
$
f~\rightarrow~-{\rm e}^{i\pi\alpha(f)}f~,~~\alpha(f)\in(-1,+1].
$
Each set of specified
phases for all world-sheet fermions, around all the non-contractible
loops is called the spin structure of the model. Such spin structures
are usually given in the form of 64 dimensional boundary condition vectors,
with each element of the vector specifying the phase of the corresponding
world-sheet fermion. The basis vectors are constrained by string consistency
requirements and completely determine the vacuum structure of the model.
The physical spectrum is obtained by applying the generalized GSO
projections
\cite{fff}.

The boundary condition basis defining a typical realistic free
fermionic heterotic string model is constructed in two stages. The
first stage consists of the NAHE set, which is a set of five
boundary condition basis vectors, $\{ 1 ,S,b_1,b_2,b_3\}$
\cite{costas,nahe}. The gauge group induced by the NAHE set is
${\rm SO} (10)\times {\rm SO}(6)^3\times {\rm E}_8$ with ${ N}=1$
supersymmetry. The space-time vector bosons that generate the
gauge group arise from the Neveu--Schwarz sector and from the
sector $\xi_2\equiv 1+b_1+b_2+b_3$. The Neveu-Schwarz sector
produces the generators of ${\rm SO}(10)\times {\rm SO}(6)^3\times
{\rm SO}(16)$. The $\xi_2$-sector produces the spinorial 128 of
SO(16) and completes the hidden gauge group to ${\rm E}_8$. The
NAHE set divides the internal world-sheet fermions in the
following way: ${\bar\phi}^{1,\cdots,8}$ generate the hidden ${\rm
E}_8$ gauge group, ${\bar\psi}^{1,\cdots,5}$ generate the SO(10)
gauge group, and $\{{\bar y}^{3,\cdots,6},{\bar\eta}^1\}$,
$\{{\bar y}^1,{\bar
y}^2,{\bar\omega}^5,{\bar\omega}^6,{\bar\eta}^2\}$,
$\{{\bar\omega}^{1,\cdots,4},{\bar\eta}^3\}$ generate the three
horizontal ${\rm SO}(6)$ symmetries. The left-moving
$\{y,\omega\}$ states are divided into $\{{y}^{3,\cdots,6}\}$,
$\{{y}^1,{y}^2,{\omega}^5,{\omega}^6\}$,
$\{{\omega}^{1,\cdots,4}\}$ and $\chi^{12}$, $\chi^{34}$,
$\chi^{56}$ generate the left-moving ${ N}=2$ world-sheet
supersymmetry. At the level of the NAHE set the sectors $b_1$,
$b_2$ and $b_3$ produce 48 multiplets, 16 from each, in the $16$
representation of SO(10). The states from the sectors $b_j$ are
singlets of the hidden ${\rm E}_8$ gauge group and transform under
the horizontal ${\rm SO}(6)_j$ $(j=1,2,3)$ symmetries. This
structure is common to all known realistic free fermionic models.

The second stage of the
construction consists of adding to the
NAHE set three (or four) additional basis vectors.
These additional vectors reduce the number of generations
to three, one from each of the sectors $b_1$,
$b_2$ and $b_3$, and simultaneously break the four dimensional
gauge group. The assignment of boundary conditions to
$\{{\bar\psi}^{1,\cdots,5}\}$ breaks SO(10) to one of its subgroups
${\rm SU}(5)\times {\rm U}(1)$ \cite{fsu5}, ${\rm SO}(6)\times {\rm SO}(4)$
\cite{alr},
${\rm SU}(3)\times {\rm SU}(2)\times {\rm U}(1)^2$ \cite{fny,euslm,cfn},
${\rm SU}(3)\times {\rm SU}(2)^2\times {\rm U}(1)$ \cite{cfs} or
${\rm SU}(4)\times {\rm SU}(2)\times {\rm U}(1)$ \cite{cfnooij}.
Similarly, the hidden ${\rm E}_8$ symmetry is broken to one of its
subgroups, and the flavor ${\rm SO}(6)^3$ symmetries are broken
to $U(1)^n$, with $3\le n\le9$.
For details and phenomenological studies of
these three generation string models we refer interested
readers to the original literature and review articles
\cite{reviewsp}.

The correspondence of the free fermionic models
with the orbifold construction is illustrated
by extending the NAHE set, $\{ 1,S,b_1,b_2,b_3\}$, by at least
one additional boundary condition basis vector \cite{foc}
\beq
\xi_1=(0,\cdots,0\vert{\underbrace{1,\cdots,1}_{{\bar\psi^{1,\cdots,5}},
{\bar\eta^{1,2,3}}}},0,\cdots,0)~.
\label{vectorx}
\eeq
With a suitable choice of the GSO projection coefficients the
model possesses an ${\rm SO}(4)^3\times {\rm E}_6\times {\rm U}(1)^2
\times {\rm E}_8$ gauge group
and ${ N}=1$ space-time supersymmetry. The matter fields
include 24 generations in the 27 representation of
${\rm E}_6$, eight from each of the sectors $b_1\oplus b_1+\xi_1$,
$b_2\oplus b_2+\xi_1$ and $b_3\oplus b_3+\xi_1$.
Three additional 27 and $\overline{27}$ pairs are obtained
from the Neveu-Schwarz $\oplus~\xi_1$ sector.

To construct the model in the orbifold formulation one starts
with the compactification on a torus with nontrivial background
fields \cite{Narain}.
The subset of basis vectors
\beq
\{ 1,S,\xi_1,\xi_2\}
\label{neq4set}
\eeq
generates a toroidally-compactified model with ${ N}=4$ space-time
supersymmetry and ${\rm SO}(12)\times {\rm E}_8\times {\rm E}_8$ gauge group.
The same model is obtained in the geometric (bosonic) language
by tuning the background fields to the values corresponding to
the SO(12) lattice. The
metric of the six-dimensional compactified
manifold is then the Cartan matrix of SO(12),
while the antisymmetric tensor is given by
\begin{equation}\label{bso12}
B_{ij}=\left\{ \begin{matrix} G_{ij} & i>j,\\ 0 & i=j,\\ -G_{ij} & i<j.
\end{matrix} \right.
\end{equation}
When all the radii of the six-dimensional compactified
manifold are fixed at $R_I=\sqrt2$, it is seen that the
left- and right-moving momenta
$
P^I_{R,L}=[m_i-{\frac{1}{2}}(B_{ij}{\pm}G_{ij})n_j]{e_i^I}^*
$
reproduce the massless root vectors in the lattice of
SO(12). Here $e^i=\{e_i^I\}$ are six linearly-independent
vielbeins normalized so that $(e_i)^2=2$.
The ${e_i^I}^*$ are dual to the $e_i$, with
$e_i^*\cdot e_j=\delta_{ij}$.

Adding the two basis vectors $b_1$ and $b_2$ to the set
(\ref{neq4set}) corresponds to the ${Z}_2\times {Z}_2$
orbifold model with standard embedding.
Starting from the $N=4$ model with ${\rm SO}(12)\times
{\rm E}_8\times {\rm E}_8$
symmetry~\cite{Narain}, and applying the ${Z}_2\times {Z}_2$
twist on the
internal coordinates, reproduces
the spectrum of the free-fermion model
with the six-dimensional basis set
$\{ 1,S,\xi_1,\xi_2,b_1,b_2\}$.
The Euler characteristic of this model is 48 with $h_{11}=27$ and
$h_{21}=3$.

It is noted that the effect of the additional basis vector $\xi_1$ of eq.
(\ref{vectorx}), is to separate the gauge degrees of freedom, spanned by
the world-sheet fermions $\{{\bar\psi}^{1,\cdots,5},
{\bar\eta}^{1},{\bar\eta}^{2},{\bar\eta}^{3},{\bar\phi}^{1,\cdots,8}\}$,
from the internal compactified degrees of freedom $\{y,\omega\vert
{\bar y},{\bar\omega}\}^{1,\cdots,6}$.
In the realistic free fermionic
models this is achieved by the vector $2\gamma$ \cite{foc}, with
\beq
2\gamma=(0,\cdots,0\vert{\underbrace{1,\cdots,1}_{{\bar\psi^{1,\cdots,5}},
{\bar\eta^{1,2,3}} {\bar\phi}^{1,\cdots,4}} },0,\cdots,0)~,
\label{vector2gamma}
\eeq
which breaks the ${\rm E}_8\times {\rm E}_8$ symmetry to ${\rm SO}(16)\times
{\rm SO}(16)$.
The ${Z}_2\times {Z}_2$ twist induced by $b_1$ and $b_2$
breaks the gauge symmetry to
${\rm SO}(4)^3\times {\rm SO}(10)\times {\rm U}(1)^3\times {\rm SO}(16)$.
The orbifold still yields a model with 24 generations,
eight from each twisted sector,
but now the generations are in the chiral 16 representation
of SO(10), rather than in the 27 of ${\rm E}_6$. The same model can
be realized with the set
$\{ 1,S,\xi_1,\xi_2,b_1,b_2\}$,
by projecting out the $16\oplus{\overline{16}}$
from the $\xi_1$-sector taking
\beq\label{changec}
\cc{\xi_1}{\xi_2} \rightarrow -\cc{\xi_1}{\xi_2}.
\eeq
This choice also projects out the massless vector bosons in the
128 of SO(16) in the hidden-sector ${\rm E}_8$ gauge group, thereby
breaking the ${\rm E}_6\times {\rm E}_8$ symmetry to
${\rm SO}(10)\times {\rm U}(1)\times {\rm SO}(16)$.
We can define two ${ N}=4$ models generated by the set
(\ref{neq4set}), ${ Z}_+$ and ${ Z}_-$, depending on the sign
in eq. (\ref{changec}). The first, say ${ Z}_+$,
produces the ${\rm E}_8\times {\rm E}_8$ model, whereas the second, say
${ Z}_-$, produces the ${\rm SO}(16)\times {\rm SO}(16)$ model.
However, the ${Z}_2\times
{Z}_2$
twist acts identically in the two models, and their physical characteristics
differ only due to the discrete torsion eq. (\ref{changec}).

This analysis confirms that the $Z_2\times Z_2$
orbifold on the $SO(12)$ lattice is at the core of the realistic
free fermionic models.
To illustrate how the chiral generations are generated in the free
fermionic models we consider the $E_6$ model which is produced
by the extended NAHE--set $\{ 1,S,\xi_1,\xi_2,b_1,b_2\}$.

The chirality of the states from a twisted sector
$b_j$ is determined by the free phase $\cc{b_j}{b_i}$.
Since we have a freedom in the choice of the sign of this
free phase, we can get from the sector $(b_i)$ either the
27 or the $\overline{27}$. Which of those we obtain in the physical
spectrum depends on the sign of the free phase.
The free phases $\cc{b_j}{b_i}$ also fix the total number
of chiral generations. Since there are two $b_i$ projections
for each sector $b_j$, $i\ne j$ we can use one projections to project
out the states with one chirality and the other projection to
project out the states with the other chirality. Thus, the total
number of generations with this set of basis vectors
is given by
$$8\left(\frac{\cc{b_1}{b_2} + \cc{b_1}{b_3}}{2}\right) +
8\left(\frac{\cc{b_2}{b_1} + \cc{b_2}{b_3}}{2}\right) +
8\left(\frac{\cc{b_3}{b_1} + \cc{b_3}{b_1}}{2}\right)$$
Since the modular invariance rules fix
$\cc{b_j}{b_i}=\cc{b_i}{b_j}$ we get that
the total number of generations is either
24 or 8. Thus, to reduce the number of generation further
it is necessary to introduce additional basis vectors.

To illustrate the reduction to three generations in
the realistic free fermionic models we consider the model in table
\ref{m278}

\beqn
 &\begin{tabular}{c|c|ccc|c|ccc|c}
 ~ & $\psi^\mu$ & $\chi^{12}$ & $\chi^{34}$ & $\chi^{56}$ &
        $\bar{\psi}^{1,...,5} $ &
        $\bar{\eta}^1 $&
        $\bar{\eta}^2 $&
        $\bar{\eta}^3 $&
        $\bar{\phi}^{1,...,8} $\\
\hline
\hline
  ${\alpha}$  &  0 & 0&0&0 & 1~1~1~0~0 & 0 & 0 & 0 & 1~1~1~1~0~0~0~0 \\
  ${\beta}$   &  0 & 0&0&0 & 1~1~1~0~0 & 0 & 0 & 0 & 1~1~1~1~0~0~0~0 \\
  ${\gamma}$  &  0 & 0&0&0 &
        $\frac{1}{2}$~$\frac{1}{2}$~$\frac{1}{2}$~$\frac{1}{2}$~$\frac{1}{2}$
          & $\frac{1}{2}$ & $\frac{1}{2}$ & $\frac{1}{2}$ &
                $\frac{1}{2}$~0~1~1~$\frac{1}{2}$~$\frac{1}{2}$~$\frac{1}{2}$~0
\\
\end{tabular}
   \nonumber\\
   ~  &  ~ \nonumber\\
   ~  &  ~ \nonumber\\
     &\begin{tabular}{c|c|c|c}
 ~&   $y^3{y}^6$
      $y^4{\bar y}^4$
      $y^5{\bar y}^5$
      ${\bar y}^3{\bar y}^6$
  &   $y^1{\omega}^5$
      $y^2{\bar y}^2$
      $\omega^6{\bar\omega}^6$
      ${\bar y}^1{\bar\omega}^5$
  &   $\omega^2{\omega}^4$
      $\omega^1{\bar\omega}^1$
      $\omega^3{\bar\omega}^3$
      ${\bar\omega}^2{\bar\omega}^4$ \\
\hline
\hline
$\alpha$ & 1 ~~~ 0 ~~~ 0 ~~~ 0  & 0 ~~~ 0 ~~~ 1 ~~~ 1  & 0 ~~~ 0 ~~~ 1 ~~~ 1
\\
$\beta$  & 0 ~~~ 0 ~~~ 1 ~~~ 1  & 1 ~~~ 0 ~~~ 0 ~~~ 0  & 0 ~~~ 1 ~~~ 0 ~~~ 1
\\
$\gamma$ & 0 ~~~ 1 ~~~ 0 ~~~ 1  & 0 ~~~ 1 ~~~ 0 ~~~ 1  & 1 ~~~ 0 ~~~ 0 ~~~ 0
\\
\end{tabular}
\label{m278}
\eeqn

Here the vector $\xi_1$ (\ref{vectorx})
is replaced by the vector $2\gamma$ (\ref{vector2gamma}).
At the level of the NAHE set we have 48 generations.
One half of the generations is projected by the vector $2\gamma$.
Each of the three vectors in table \ref{m278}
acts nontrivially on the degenerate
vacuum
of the sectors $b_1$, $b_2$ and $b_3$ and reduces the
number of generations in each step
by a half. Thus, we obtain one generation from each sector
$b_1$, $b_2$ and $b_3$.

The geometrical interpretation of the basis vectors
beyond the NAHE set is facilitated by taking combinations of the
basis vectors in \ref{m278}, which entails choosing another set
to generate the same vacuum. The combinations
$\alpha+\beta$, $\alpha+\gamma$, $\alpha+\beta+\gamma$ produce
the following boundary conditions under the set of internal
real fermions

\begin{eqnarray}
     &\begin{tabular}{c|c|c|c}
 ~&   $y^3{y}^6$
      $y^4{\bar y}^4$
      $y^5{\bar y}^5$
      ${\bar y}^3{\bar y}^6$
  &   $y^1{\omega}^5$
      $y^2{\bar y}^2$
      $\omega^6{\bar\omega}^6$
      ${\bar y}^1{\bar\omega}^5$
  &   $\omega^2{\omega}^4$
      $\omega^1{\bar\omega}^1$
      $\omega^3{\bar\omega}^3$
      ${\bar\omega}^2{\bar\omega}^4$ \\
\hline
\hline
$\alpha+\beta$
& 1 ~~~ 0 ~~~ 1 ~~~ 1  & 1 ~~~ 0 ~~~ 1 ~~~ 1  & 0 ~~~ 1 ~~~ 1 ~~~ 0
\\
$\beta+\gamma$
& 0 ~~~ 1 ~~~ 1 ~~~ 0  & 1 ~~~ 1 ~~~ 0 ~~~ 1  & 1 ~~~ 1 ~~~ 0 ~~~ 1
\\
$\alpha+\beta+\gamma$
& 1 ~~~ 1 ~~~ 1 ~~~ 0  & 1 ~~~ 1 ~~~ 1 ~~~ 0  & 1 ~~~ 1 ~~~ 1 ~~~ 0
\\
\\
\end{tabular}
\label{m2782}
\end{eqnarray}

It is noted that the two combinations $\alpha+\beta$ and $\beta+\gamma$
are fully symmetric between the left and right movers, whereas the
third, $\alpha+\beta+\gamma$, is asymmetric.
The action of the first two combinations
on the compactified bosonic coordinates translates therefore to symmetric
shifts. Thus, we see that reduction of the number of generations
is obtained by further action of symmetric shifts.

Due to the presence of the third combination the situation, however,
is more complicated. The third combination in (\ref{m2782}) is
asymmetric between the left and right movers and therefore
does not have an obvious geometrical interpretation.
Below we perform a complete
classification of all the possible NAHE--based
$Z_2\times Z_2$ orbifold models with symmetric shifts on the complex
tori, which reveals that three generations are not obtained in this manner.
Three generations are obtained in the free fermionic models
by the inclusion of the asymmetric shift in (\ref{m2782}).
This outcome has profound implications on the type of geometries that
may be related to the realistic string vacua, as well as on the
issue of moduli stabilization.

\section{$N=1$ heterotic orbifold constructions}\label{three}

In this section we revise the $Z_2\times Z_2$
heterotic orbifold construction and relate this to
 the free fermionic construction. We isolate the individual conformal blocks
that will facilitate the classification of the models and set up a procedure
to analyse all possible $N=1$ heterotic $Z_2 \times Z_2 $ models. We start by
describing the procedure to descend from $N=4$ to $N=1$ supersymmetric
heterotic vacua.

\subsection{The $N=4$ models}

The partition function for any  heterotic model via the fermionic construction is
\begin{equation}\label{eq:fermionpt}
Z = \frac{1}{\tau_2} \frac{1}{\eta^{12}\bar{\eta}^{24}} \sum_{a,b \in \Xi} c[^a
_b] \frac{1}{2^{M}} \prod_{i=1}^{20} \theta[^{a_i}_{b_i}]^{\frac{1}{2}} \prod
_{j=1}^{44} \bar{\theta}[^{a_j}_{b_j}]^{\frac{1}{2}}.
\end{equation}
In the above  equation $M$ is the number of basis vectors and the
parameters in the $\theta$--functions represent the action of
the vectors. In order to obtain a supersymmetric model we need at least two
basis vectors $\{1,S\}$.
\begin{eqnarray}
1 &=& \{ \psi^{1,2}, \chi^{1,\ldots,6}, y^{1,\ldots,6}, \omega^{1,\ldots,6} |
\nonumber\\
  & & \bar{y}^{1,\ldots,6}, \bar{\omega}^{1,\ldots,6}, \bar{\psi}^{1,\ldots,6},
 \bar{\eta}^{1,2,3}, \bar{\phi}^{1,\ldots,8} \} ,\label{eq:1}\\
S &=& \{ \psi^{1,2}, \chi^{1,\ldots,6} \} .\label{eq:S}
\end{eqnarray}
The supersymmetric GSO projection is induced by the set $S$ for any choice
of the  GSO coefficient
\begin{equation}
\cc{S}{1}=\pm1.
\end{equation}
The corresponding partition function has a factorized left--moving
contribution coming from the sector $S$,
\begin{equation}\label{eq:fermionpt44}
Z_{1,S} = \frac{1}{\tau_2 |\eta |^4 } \frac{1}{2} \sum_{a,b=0}^1
(-1)^{a+b+\mu ab} \frac{\theta[^a_b]^4}{\eta^4}
\frac{\Gamma_{6,6+16}[SO(44)]}{\eta^6\bar{\eta}^{22}}
\end{equation}
where
\begin{equation}
{\Gamma_{6,6+16}[SO(44)]}=\frac{1}{2} \sum_{c,d}
\frac{\theta[^c_d]^6 \bar{\theta}[^c_d]^{22}}{\eta^6\bar{\eta}^{22}},
\end{equation}
and
$$ \mu=\frac{1}{2}{\left(1-\cc{S}{1}\right)}$$ defines the chirality of $N=4$
supersymmetry. Therefore, the role of the boundary condition vector $S$ is to
factorize the left--moving contribution,
\begin{equation}\label{eq:Jacobi}
Z^L_{N=4}= \frac{1}{2 } \sum_{a,b=0}^1
(-1)^{a+b+\mu ab}\theta[^a_b](v)\theta[^a_b]^3(0)\sim v^4
\end{equation}
which is zero with the multiplicity of $N=4$ supersymmetry.

The above partition function gives rise to an $SO(44)$ right--moving
gauge group and is the maximally symmetric point in the moduli space of
the Narain $\Gamma_{6,6+16}$ lattice. The  general $\Gamma_{6,6+16}$ lattice
depends on $6\times 22$ moduli, the metric $G_{ij}$ and the antisymmetric
tensor $B_{ij}$  of the six dimensional internal space, as well as the
Wilson lines $Y^I_i$ that appear in the  2d-world--sheet.

\begin{eqnarray}
S &=& \frac{1}{4\pi} \int d^2\sigma \sqrt{g}g^{ab}G_{ij}\partial_a
X^i\partial_bX^j + \frac{1}{4\pi} \int d^2\sigma \epsilon^{ab}
B_{ij}\partial_aX^i\partial_bX^j\nonumber\\
  & & + \frac{1}{4\pi} \int d^2\sigma \sqrt{g} \sum_I \psi^I \Big[ \bar{\nabla}
 + Y^I_i\bar{\nabla}X^i \Big]\bar{\psi}^I.
\end{eqnarray}
Here $i$ runs over the internal coordinates and $I$ runs over the extra $16$
right--moving degrees of freedom described by $\bar{\psi}^I$.

The compactified sector of the partition function is given by
$\Gamma_{6,6+16}$
\begin{eqnarray}\label{eq:internal}
\Gamma_{6,6+16} &=& \frac{(\det G)^3}{\tau_2^3}\sum_{m,n} \exp\bigg\{ -\pi
\frac{T_{ij}}{\tau_2}[m^i + n^i\tau][m^j + n^j\bar{\tau}]\bigg\}\\
& & \times \frac{1}{2}\sum_{\gamma,\delta} \prod_{I=1}^{16}\exp\Big[-i\pi n^i
\big(m^j + n^j \bar{\tau}\big)Y^I_iY^I_j\Big]
\bartheta{\gamma}{\delta}\big(Y^I_i
(m^i+n^i\bar{\tau})|\tau\big),\nonumber
\end{eqnarray}
where $T_{ij} = G_{ij} + B_{ij}$.

Equation \eqref{eq:internal} is the winding mode representation of the
partition function. Using a
Poisson resummation we can put it  in the momentum representation form:
\begin{equation}
\Gamma_{6,22} = \sum_{P, \bar{P}, Q} \exp\bigg\{\frac{i\pi\tau}{2}P_iG^{ij}P_j
-\frac{i\pi\bar{\tau}}{2}\bar{P}_iG^{ij}\bar{P}_j - i\pi\bar{\tau}\hat{Q}^I\hat
{Q}^I\bigg\},
\end{equation}
with
\begin{eqnarray}
P_i &=& m_i +B_{ij}n^j + \frac{1}{2}Y^I_iY^I_jn^j + Y^I_iQ^I + G_{ij}n^j\\
\bar{P}_i &=& m_i +B_{ij}n^j + \frac{1}{2}Y^I_iY^I_jn^j + Y^I_iQ^I - G_{ij}
n^j\\
\hat{Q}^I &=& Q^I + Y^I_in^i.
\end{eqnarray}
The charge momenta $Q^I$ are induced by the right--moving fermions
$\bar{\psi}^I$ which  appear explicitly in the $\theta$--functions.
\begin{equation}
\theta[^{a^I}_{b^I}] = \sum_{n\in Z} q^{\frac{(Q^I)^2}{2}} e^{2\pi i
(v-{\frac{b^I}{2}}) Q^I},
\end{equation}
where the charge momentum  $Q^I=(n-{\frac{a^I}{2}})$.

For generic $G_{ij}, B_{ij}$ and for vanishing values for  Wilson lines,
$Y^I_i=0$
one  obtains an $N=4$ model with a gauge group $U(1)^6 \times SO(32)$.
The $U(1)^6$ can be extended to $SO(12)$ by fixing the moduli of the
internal manifold \cite{foc}.

The $N=4$ fermionic construction based on $\{1,S\}$\eqref{eq:fermionpt44}
has an  extended  gauge group, $SO(44)$. From the lattice construction
point of view, an $N=4$ model with a gauge group $G \subset SO(44)$
can be generated by switching on Wilson lines and fine tune the moduli
of the internal manifold.
Moving from the $SO(44)$ to  $U(1)^6 \times SO(32)$ heterotic point as well
as to the  $U(1)^6 \times E_8 \times E_8$ point can be realized continuously
\cite{Kiritsis:1997ca}.
The partition function at the $U(1)^6 \times E_8 \times E_8$ point takes
a simple factorized form
\begin{eqnarray}
\Gamma_{6,6+16} &=& \frac{(\det G)^3}{\tau_2^3}\sum_{m,n} \exp\bigg\{ -\pi
\frac{T_{ij}}{\tau_2}[m^i + n^i\tau][m^j + n^j\bar{\tau}]\bigg\}
\label{eq:internal2}\\
& & \times \frac{1}{2}\sum_{\gamma,\delta} \bartheta{\gamma}{\delta}^8
\times  \frac{1}{2}\sum_{\gamma,\delta} \bartheta{\gamma+h}{\delta+g}^8
\nonumber
\end{eqnarray}

\subsection{The $N=1$ models}\label{sec:n1}

To break the number of supersymmetries down from $N=4$ to $N=1$ in the
fermionic formulation we need to introduce the vectors $b_1$ and $b_2$.
\begin{eqnarray}
b_1 &=& \{ \chi^{3,4}, \chi^{5,6},\ y^{3,4},y^{5,6}\ |\ \ldots\  \} ,
\label{eq:b1flat}\\
b_2 &=& \{ \chi^{1,2}, \chi^{5,6}, y^{1,2},\ y^{5,6}\ |\ \ldots\ \} .
\label{eq:b2flat}
\end{eqnarray}
The $ b_1$ twists the second  and third complex planes (3,4)and (5,6) while
$ b_2$ twists the first  and third (1,2) and (5,6) ones. Thus, $b_1,b_2$
separate the internal lattice into the three complex  planes:
(1,2), (3,4) and (5,6).

The action of the $ b_i$--twists  fully determines the fermionic
content for the left--moving sector. The dots $ \ \ldots\ $
in $ b_1$, $ b_2$ stand for the $n_1$, $ n_2$ right--moving fermions.
To generate a modular invariant model we can distinguish four options.
$n_i$ are either $8$, $16$, $24$ or $32$ real right-moving fermions in
the basis vector $b_i$.

Defining the basis vectors with $8$ real right--moving fermions leads to
massless states in the spectrum in  vectorial representations of the gauge
groups; $16$ real right--moving fermions give rise to
spinorial representations on each plane. Adding either $24$ or $32$
right--moving  fermions would produce massive states in the spectrum.
We therefore discard the last two options. We thus need to introduce
$16$ real fermions ($8$ complex) in the vectors $b_1,b_2$ for the
existence of spinorial representations on the first and second plane.

A suitable choice is for instance,
\begin{eqnarray}
b_1 &=& \{ \chi^{3,4}, \chi^{5,6}, y^{3,4}, y^{5,6}\
| \bar{y}^{3,4}, \bar{y}^{5,6},\
\bar{\eta}^{1}, \bar{\psi}^{1,\ldots,5} \} ,\label{eq:twistvectors1}\\
b_2 &=& \{ \chi^{1,2}, \chi^{5,6}, y^{1,2}, y^{5,6}\
| \bar{y}^{1,2}, \bar{y}^{5,6},\
\bar{\eta}^{2}, \bar{\psi}^{1,\ldots,5} \} .
\label{eq:twistvectors2}
\end{eqnarray}
We define the vectors $x=\{0,..., | \bar{\psi}^{1,\ldots,5} ,
\bar{\eta}^{1,2,3}\}$, and ${\tilde b}_{1,2}=b_{1,2}+x$.
The  $N=1$ partition function based on
$\{1,S,{\tilde b}_1,{\tilde b}_2\}$
takes the following form:

\begin{eqnarray}
Z_{N=1} &=&
\frac{1}{\tau_2 |\eta |^4 } \frac{1}{2}
\sum_{\alpha,\beta} e^{i\pi(a+b+\mu ab)}
\frac{1}{4}\sum_{h_1,h_2,g_1,g_2}\  {\frac{\theta[^a_b]}{\eta}}\
{ \frac{\theta[^{a+h_2}_{b+g_2}]}{\eta}}\
 {\frac{\theta[^{a+h_1}_{b+g_1}]}{\eta}}\
{\frac{\theta[^{a-h_1-h_2}_{b-g_1-g_2}]}{\eta}} \nonumber\\
&  & \times  \frac{1}{2}\sum_{\gamma,\delta}
 {{ \frac{\Gamma_{6,6}\left[^{\gamma,h_1,h_2}_{\delta,g_1,g_2}\right]}{
{\eta^6 {\bar\eta}^6}}} \times {\frac{Z_{\eta}\left[^{\gamma,h_1,h_2}_{\delta,g_1,g_2}\right]}{
{\bar \eta}^3}}
\times  \frac{Z_{26}\left[^{\gamma}_{\delta} \right]}{\bar
\eta}^{13}}\ e^{i\pi\varphi_L}
\label{eq:partition}\\
\Gamma_{6,6} \left[^{\gamma,h_1,h_2}_{\delta,g_1,g_2}\right]&=&
\Big|\theta[^{\gamma}_{\delta}]\theta[^{\gamma+h_2}_{\delta+g_2}] \Big|^2
\Big|\theta[^{\gamma}_{\delta}]\theta[^{\gamma+h_1}_{\delta+g_1}]\Big|^2\
\Big|\theta[^{\gamma}_{\delta}]\theta[^{\gamma-h_1-h_2}_{\delta-g_1-g_2}]
\Big|^2
\label{eq:selfdual}\\
Z_{\eta}\left[^{\gamma,h_1,h_2}_{\delta,g_1,g_2}\right] &=& \bar{\theta}[^{\gamma+h_2}_{\delta+g_2}]
\bar{\theta}[^{\gamma+h_1}_{\delta+h_2}]\
\bar{\theta}[^{\gamma-h_1-h_2}_{\delta-g_1-g_2}]\\
Z_{26}\left[^{\gamma}_{\delta} \right] &=& \bar{\theta}[^{\gamma}_{\delta}]^{13}\label{eq:gaugesector}
\end{eqnarray}

In equation \eqref{eq:selfdual} the internal manifold is twisted and
thereby separated explicitly into three planes.
The above model is the minimal  $Z_2 \times Z_2$ with  $N=1$ supersymmetry
and massless spinorial representations in the same $SO(10)$ group coming
from the first and/or from the second plane. The number of families
depends on the choice of the phase $\varphi_L$.
The freedom of this phase arises from the different possible choices
of the modular invariance coefficients  $c[^{v_i}_{v_j}]$.
The maximal number of the families for this model is $32$.
Introducing internal  shifts, associated to  $\varphi_L$,
can reduce this number as we will discuss below.

We could have chosen the boundary conditions for different right-moving
fermions. This would lead to spinorial representations on each plane, but the
group to which they would belong would differ in each plane. As we require
spinors in the same group we have discarded this option. Choosing an overlap
with more than $6$ complex fermions in the right-moving sector between the
vectors $b_1$ and $b_2$ leads to a $SO(14)$ gauge group, which does not
have  chiral fermions.

In order to have spinors in the spectrum on \emph{all} three planes we need to
separate at least an $SO(16)$ (or $E_8$) from the $\Gamma_{6,22}$ lattice . We
therefore need to introduce the additional vector
\begin{equation}\label{eq:z}
z = \{ \bar{\phi}^{1,\ldots,8} \}
\end{equation}
to the set. With this vector the partition function for the gauge sector
\eqref{eq:gaugesector} modifies to
\begin{equation}\label{eq:gamma013}
Z_{26}\left[^{\gamma,h_z}_{\delta,g_z} \right] =
 \frac{1}{2} \sum_{h_z,g_z} \bar{\theta}[^{\gamma}_{\delta}]^{5}
\bar{\theta}[^{\gamma+h_z}_{\delta+g_z}]^{8}.
\end{equation}
We can further  separate out the internal  $\Gamma_{6,6}$ lattice by
introducing the additional vector,
\begin{equation}\label{eq:internalseparation}
e = \{ y_{1,\ldots,6}, \omega_{1,\ldots,6}
| \bar{y}_{1,\ldots,6}, \bar{\omega}_{1,\ldots,6} \} ,
\end{equation}
which modifies the  $\Gamma_{6,6}$ in \eqref{eq:selfdual} by
\begin{equation}\label{eq:gamma6,6}
\Gamma_{6,6} \left[^{\gamma,h_1,h_2}_{\delta,g_1,g_2}\right]=
 \frac{1}{2} \sum_{h_e,g_e}
\Big|\theta[^{\gamma+h_e}_{\delta+g_e}]
\theta[^{\gamma+h_e+h_2}_{\delta+g_e+g_2}] \Big|^2
\Big|\theta[^{\gamma+h_e}_{\delta+g_e}]
\theta[^{\gamma+h_e+h_1}_{\delta+g_e+g_1}]\Big|^2
\Big|\theta[^{\gamma+h_e}_{\delta+g_e}]
\theta[^{\gamma+h_e-h_1-h_2}_{\delta+g_e-g_1-g_2}]\Big|^2.
\end{equation}
In the above $\{1,S,b_1,b_2,e\}$ construction the gauge group of the
observable sector becomes either $SO(10) \times U(1)^3$
or $E_6 \times U(1)^2$ and the hidden sector necessarily is $SO(16)$ or $E_8$
depending on the generalized GSO coefficients, (the choice of the phase
$\varphi_L$), while the gauge group from the $\Gamma_{6,6}$ lattice becomes
$G_L=SO(6)\times U(1)^3$.

So far the construction of the $N=1$ models is generic. The only
requirement we are imposing is the presence of
{\em spinors on all three planes}. We call this the $S^3$ subclass of models.
In a general $N=1$ model the spinors could be replaced by vectorial
representations of the observable gauge group. This replacement gives rise
to three additional classes of models which we denote by  $S^2V$, $SV^2$
and $V^3$. In this work we will focus on the $S^3$ class  and we will deal
with the other  classes in a future work.
The condition of spinorial representations arising from
each one of the $Z_2 \times Z_2$ orbifold planes together with the complete
separation of the internal manifold is synonymous to having a well defined
hidden gauge group.

\subsection{The general $S^3$ $N=1$ model}\label{sec:n1general}

In the class of $Z_2 \times Z_2$ orbifold models, the internal manifold is
broken into three planes. The hidden gauge group is necessarily $E_8$ or
$SO(16)$  broken to any subgroup by Wilson lines (at the $N=4$ level).
In order to  classify all possible $S^3$ models, it is necessary to
consider all possible basis vectors consistent with modular invariance.
Namely:
\begin{eqnarray}
z_1 &=& \{ \bar{\phi}^{1,\ldots,4} \}  \label{eq:z1},\\
z_2 &=& \{ \bar{\phi}^{5,\ldots,8} \}  \label{eq:z2},\\
e_i &=& \{ y_i, \omega_i | \bar{y}_i, \bar{\omega}_i \} ,\
\ i \in \{1,2,3,4,5,6\}.
\end{eqnarray}

The $z_1,z_2$ vectors allow for a breaking of hidden $E_8$ or $SO(16)$ to
$SO(8) \times SO(8)$ depending on the modular  coefficients.
As we discuss below this splitting of the hidden
gauge group has important consequences in the
classification of the $S^3$ class of models by the number of generations.
The introduction of $e_i$ vectors is  necessary in order to obtain all possible
internal shifts which also induces all possible  modification to the number
of generations.

The general $N=1$, $S^3$ model based on
$\{1,S,e_i,z_1,z_2,{\tilde b}_1,{\tilde b}_2\}$ is:
\begin{eqnarray}
Z_{N=1} &=&
\frac{1}{\tau_2 |\eta |^4 } \frac{1}{2}
\sum_{\alpha,\beta} e^{i\pi(a+b+\mu ab)}
\frac{1}{4}\sum_{h_1,h_2,g_1,g_2}\  {\frac{\theta[^a_b]}{\eta}}\
{\frac{\theta[^{a+h_1}_{b+g_1}]}{\eta}}\
{\frac{\theta[^{a+h_2}_{b+g_2}]}{\eta}}\
{\frac{\theta[^{a-h_1-h_2}_{b-g_1-g_2}]}{\eta}} \nonumber\\
&  & \times \frac{1}{2^6}\sum_{p_i,q_i}
\frac{\Gamma_{2,2} \left[^{h_1|p_1,p_2}_{g_1|q_1,q_2} \right]}{\eta^2 {\bar\eta}^2}
\frac{\Gamma_{2,2} \left[^{h_2|p_3,p_4}_{g_2|q_3,q_4} \right]}{\eta^2 {\bar\eta}^2}
\frac{\Gamma_{2,2} \left[^{-h_1-h_2|p_5,p_6}_{-g_1-g_2|q_5,q_6} \right]}{\eta^2 {\bar\eta}^2}
\nonumber\\
&  & \times \frac{1}{8}\sum_{\gamma,\gamma',\xi,\delta,\delta',\zeta}
\frac{Z_{\eta}\left[^{\gamma,h_1,h_2}_{\delta,g_1,g_2} \right]}{{\bar \eta}^3}
\times \frac{Z_{10}\left[^{\gamma}_{\delta} \right]}{{\bar\eta}^{5}}
\times \frac{Z_{16}\left[^{\gamma',\xi}_{\delta', \zeta} \right]}{{\bar\eta}^{8}} \ e^{i\pi\varphi_L}
\label{eq:N1partition}\\
Z_{\eta}\left[^{\gamma,h_1,h_2}_{\delta,g_1,g_2}\right] &=&
\bar{\theta}[^{\gamma+h_2}_{\delta+g_2}]
\bar{\theta}[^{\gamma+h_1}_{\delta+h_2}]\
\bar{\theta}[^{\gamma-h_1-h_2}_{\delta-g_1-g_2}]\\
Z_{10}\left[^{\gamma}_{\delta} \right] &=&
\bar{\theta}[^{\gamma}_{\delta}]^{5}\\
Z_{16}\left[^{\gamma',\xi}_{\delta',\zeta} \right] &=&
\bar{\theta}[^{\gamma'}_{\delta'}]^{4}
\bar{\theta}[^{\gamma'+\xi}_{\delta'+\zeta}]^{4}
\end{eqnarray}
The $\Gamma_{6,6}$ lattice of $N=4$ is twisted by $h_i,g_i$, thus in the
$N=1$ case separated into three (2,2) planes. The contribution of each
of these planes in $N=1$ partition function is written in terms of twisted
by $h_i,g_i$ and shifted by $p_i, q_i$  $\Gamma_{2,2}$ lattice. The
expressions of those lattices at the self dual point (fermionic
construction point) is:
\begin{equation}\label{eq:fermpoint}
\Gamma_{2,2} \left[^{h|p_i,p_j}_{g|q_i,q_j} \right]|_{f.p}=
\frac{1}{4}\sum_{a_i,b_i,a_j,b_j} e^{i\pi\phi_1 +i\pi\phi_2}
\left|\theta[^{a_i}_{b_i}]\theta[^{a_i+h}_{b_i+g}]
\theta[^{a_j}_{b_j}]\theta[^{a_j+h}_{b_j+g}] \right|
\end{equation}
where the phases
$$
\phi_i=a_iq_i+b_ip_i+q_ip_i, ~~~~~~~~~\phi_j=a_jq_j+b_jp_j+q_jp_j
$$
define the two shifts of the $\Gamma_{2,2}$ lattice. At the generic point
of the moduli space the  shifted $\Gamma_{2,2}$ lattice depends on the moduli
$(T,U)$, keeping however identical modular transformation properties as
those of the fermionic point.

For non-zero twist, $(h,g)\ne (0,0)$,
$\Gamma_{2,2}$ is independent of the moduli $T,U$ and thus it is
identical to that of \eqref{eq:fermpoint} constructed at the fermionic
point\cite{Gregori:1999ny,Gregori:1999ns}.
Thus for non-- zero twist, $(h,g)\ne (0,0)$,
\begin{equation}\label{eq:twisted}
\Gamma_{2,2} \left[^{h|p_i,p_j}_{g|q_i,q_j} \right]_{(T,U)}|_{(h,g)\ne (0,0)}=
\Gamma_{2,2} \left[^{h|p_i,p_j}_{g|q_i,q_j} \right]|_{f.p}
\end{equation}

For zero twist, $(h,g)=(0,0)$, the momentum and winding modes are moduli
dependent and are shifted  by $q_i,q_j$ and $p_i, p_j$,
\begin{eqnarray}
\Gamma_{2,2} \left[^{0|p_i,p_j}_{0|q_i,q_j} \right]_{(T,U)}&&=
\sum_{\overrightarrow{m},\overrightarrow{n} \in Z} e^{i\pi \{ m_1q_i + m_2
 q_j \} }  \exp \Bigg\{ 2\pi i \bar{\tau} \left[ m_1 \left( n_1+\frac{p_i}{2}
\right) + m_2 \left( n_2+\frac{p_j}{2}\right) \right] \nonumber\\
&&-\frac{\pi \tau_2}{T_2 U_2} \left| m_1 U - m_2 + T(n_1+\frac{p_i}{2})
+ TU(n_2+\frac{p_j}{2}) \right|^2 \Bigg\}.
\end{eqnarray}
The phase  $\varphi_L$ is determined by the chirality of the supersymmetry
as well as by the other modular coefficients
\begin{equation}
\varphi_L(a,b) =  \frac{1}{2}\sum_{i,j}(1-c[^{v_i}_{v_j}])\alpha_i\beta_j,
\end{equation}
where $\alpha_i$ and $\beta_j$ are the upper-- and lower--  arguments in
$\theta$--functions corresponding to the  boundary conditions in the two
directions of the world sheet torus and which are associated to the basis
vectors $v_i$ and $v_j$ of the fermionic construction. The only freedom
which remains in the general $S^3$ $N=1$ model is therefore
the choice of the  generalized GSO projection coefficients
$c[^{v_i}_{v_j}]=\pm1$. The space of models is  classified according to that
choice  which determines at the end the phase $\varphi_L$.
We have in total $55$  independent choices for $c[^{v_i}_{v_j}]$ that can
take the values $\pm1$.
Thus, the total number of  models in this restricted class of $N=1$ models
is $2^{55}$. Latter, we will  classify all these  models according to the
values of the GSO coefficients.

The so called  NAHE models is a small sub--class of the general $S^3$, $N=1$
deformed fermionic $N=1$ model. More precisely we can write the NAHE set
basis vectors as a linear combination of
basis vectors $\{1,S,e_i,z_1,z_2,b_1,b_2\}$ which define the general $S^3$
$N=1$ model:
\begin{eqnarray}
b_1^{NAHE} &=& S + b_1,\\
b_2^{NAHE} &=& S + b_2 + e_5 + e_6, \\
b_3^{NAHE} &=& 1 + b_1 + b_2 + e_5 + e_6 + z_1 + z_2.
\end{eqnarray}
We see that the NAHE set is included in these models as mentioned in section
\ref{sec:n1}.

\subsection{The $N=4$ gauge group}

We describe the gauge configuration of the models defined by the basis
vectors $\{1,S,e_i,z_1,z_2,b_1,b_2\}$. For
this purpose we start with a simplification and separate out the internal
manifold using equation \eqref{eq:internalseparation}. As the twisting vectors
$b_1$ and $b_2$ are used to break the $SO(16) \to SO(10) \times U(1)^3$ we will
firstly describe the configuration without these vectors. The gauge group
induced by the vectors $\{1,S, e, z_1, z_2\}$ without enhancements is.
\begin{equation}
G = SO(16) \times SO(8)_1 \times SO(8)_2 \times SO(12),
\end{equation}
where the internal manifold is described by $SO(12)$ and the hidden sector by
$SO(8) \times SO(8)$ and the observable by $SO(16)$. By choosing the GSO
coefficients the SO(16) can enhance either to $E_8$ or mix with the other
sectors
producing either $SO(24)$ or $SO(32)$. Similarly the $SO(8) \times SO(8)$ can
enhance either to $SO(16)$ or $E_8$ or mix with the observable or internal
manifold gauge group. This leads to enhancements of the form $SO(20)$ or
$SO(24)$.
The exact form depends only on the three GSO coefficients $\cc{e}{z_1}$,
$\cc{e}{z_2}$, $\cc{z_1}{z_2}$. We have shown the results in table
\ref{tab:gauge}.

\begin{table}
\begin{center}
\begin{tabular}{c c c | r}
$\cc{z_1}{z_2}$ & $\cc{e}{z_1}$ & $\cc{e}{z_2}$ & Gauge group G\\\hline
+ & + & + & $E_8    \times SO(28) $\\
+ & - & + & $SO(24) \times SO(20) $\\
+ & + & - & $SO(24) \times SO(20) $\\
+ & - & - & $SO(32) \times SO(12) $\\
- & + & + & $SO(16) \times SO(16) \times SO(12)$\\
- & - & + & $SO(16) \times SO(16) \times SO(12)$\\
- & + & - & $SO(16) \times SO(16) \times SO(12)$\\
- & - & - & $E_8    \times E_8    \times SO(12)$\\
\end{tabular}
\caption{The configuration of the gauge group of the $N=4$ theory.
We have separated a priori the internal and the hidden and observable
gauge group using the vectors $e$ and $z_i$. Introducing the other
vectors $e_i$ and $b_i$ only induce breaking of these groups.}
\label{tab:gauge}
\end{center}
\end{table}

Proceeding to the complete model $\{1,S,e_i,z_1,z_2,b_1,b_2\}$
we break these gauge groups to their subgroups.
Imposing the shifts $e_i$ we can break the internal gauge group down to its
Cartan generators by a suitable choice of the coefficients. By a suitable
choice we can break $SO(20) \to SO(8) \times U(1)^6$.

When we also include the twists we break $SO(16) \to SO(10) \times U(1)^3$ and
$E_8 \to E_6 \times U(1)^2$. Similarly we can break $SO(24) \to SO(10) \times
U(1)^3 \times SO(8)$ and $SO(32) \to SO(10) \times U(1)^3 \times SO(8) \times
SO(8)$. Enhancements can subsequently occur of the form $SO(8) \times U(1)
\subset SO(32) \to SO(10)$ or $SO(8) \times SO(8) \times U(1) \subset SO(32)
\to SO(18)$. We find possible enhancements of the form $SO(10) \times SO(8)
\subset SO(32) \to SO(18)$.

In table \ref{tab:gauge} we notice that the coefficient $\cc{z_1}{z_2}$
distinguishes between the $SO(32)$ models and the $E_8 \times E_8$ models.
Since we require complete separation of the gauge group into a well defined
observable and
 hidden gauge group, we set the coefficient $\cc{z_1}{z_2}=-1$ in the
classification.



\section{Generic $Z_2\times Z_2$ model in the free fermionic formulation}
\label{four}
\newcommand{\nn}[0]{\nonumber}
\newcommand{\ba}{\begin{eqnarray}}
\newcommand{\ea}{\end{eqnarray}}
\subsection{General formalism}
In the free fermionic formulation of the heterotic superstring, a model
is determined by a set of
basis vectors, associated with the phases picked up by the fermions
when parallelly transported along
non-trivial loops and a set of coefficients associated with GSO
projections.
The free fermions in the light-cone gauge in the traditional notation
are:
$\psi^\mu, \chi^i,y^i, \omega^i, i=1,\dots,6$ (left movers)
and  $\bar{y}^i,\bar{\omega}^i, i=1,\dots,6$,
$\psi^A, A=1,\dots,5$, $\bar{\phi}^\alpha, \alpha=1,8$ (right movers).
The class of models under consideration  is generated by a set of
12 basis vectors
$$
B=\{v_1,v_2,\dots,v_{12}\}
$$
where
\begin{eqnarray}
v_1=1&=&\{\psi^\mu,\
\chi^{1,\dots,6},y^{1,\dots,6},\omega^{1,\dots,6}|\bar{y}^{1,\dots,6},
\bar{\omega}^{1,\dots,6},\bar{\eta}^{1,2,3},
\bar{\psi}^{1,\dots,5},\bar{\phi}^{1,\dots,8}\}\nn\\
v_2=S&=&\{\psi^\mu,\chi^{1,\dots,6}\}\nn\\
v_{2+i}=e_i&=&\{y^{i},\omega^{i}|\bar{y}^i,\bar{\omega}^i\}, \ i=1,\dots,6\nn\\
v_{9}=b_1&=&\{\chi^{34},\chi^{56},y^{34},y^{56}|\bar{y}^{34},\bar{y}^{56},
\bar{\eta}^1,\bar{\psi}^{1,\dots,5}\}\label{basis}\\
v_{10}=b_2&=&\{\chi^{12},\chi^{56},y^{12},y^{56}|\bar{y}^{12},\bar{y}^{56},
\bar{\eta}^2,\bar{\psi}^{1,\dots,5}\}\nn\\
v_{11}=z_1&=&\{\bar{\phi}^{1,\dots,4}\}\nn\\
v_{12}=z_2&=&\{\bar{\phi}^{5,\dots,8}\}\nn
\end{eqnarray}
The vectors $1,S$ generate an $N=4$ supersymmetric model. The vectors
$e_i,i=1,\dots,6$ give rise to
all possible symmetric shifts of internal fermions
($y^i,\omega^i,\bar{y}^i,\bar{\omega}^i$) while $b_1$ and $b_2$
stand for  the $Z_2\times Z_2$ orbifold twists. The remaining fermions
not affected by the action
of the previous vectors are $\phi^i,i=1,\dots,8$ which normally give
rise to the hidden sector gauge group.
 The vectors $z_1,z_2$ divide these eight fermions in two sets of
four which in the $Z_2\times{Z_2}$ case is the maximum consistent
partition function \cite{fff}.
This is the most general basis, with symmetric shifts for the internal
fermions, that is compatible with
Kac--Moody level one $SO(10)$ embedding.

The associated projection coefficients are denoted by
$\cc{v_i}{v_j}, i,j=1,\dots,12$ and can take the values $\pm1$.
They
are related by modular invariance
$\cc{v_i}{v_j}=\exp\{i\frac{\pi}{2}v_i\cdot v_j\} \cc{v_j}{v_i}$
and $\cc{v_i}{v_i}=\exp\{i\frac{\pi}{4}v_i\cdot v_i\} \cc{v_j}{1}$
leaving $2^{66}$ independent coefficients. Out of them, the requirement
of $N=1$ supersymmetric spectrum fixes
(up to  a phase convection) all $\cc{S}{v_i},i=1,\dots,12$. Moreover,
without loss of generality we can set $\cc{1}{1}=-1$,
and leave the rest 55 coefficients free. Therefore, a simple counting
gives  $2^{55}$ (that is approximately $10^{16.6}$) distinct models in
the class under consideration.
In the following we study this class of models by deriving  analytic
formulas for the gauge group and the
spectrum and then using these formulas for the classification.

\subsection{The gauge group}
Gauge bosons arise from the
following four sectors :
$$G=\{0,z_1,z_2,z_1+z_2,x\}$$
where
\ba
x=1+S+\sum_{i=1}^6e_i+\sum_{k=1}^2 z_k=\{{\bar{\eta}^{123},\bar{\psi}^{12345}}\}\ .
\ea
The $0$ sector gauge bosons give rise to the gauge group
$$
SO(10)\times {U(1)}^3 \times SO(8)^2
$$
The $x$ gauge bosons when present lead to enhancements of the
traditionally called observable sector (the sector that includes
$SO(10)$) while
the $z_1+z_2$ sector can enhance the hidden sector ($SO(8)^2$).
However, the $z_1, z_2$ sectors accept oscillators that can
 also give rise to mixed type gauge bosons and completely reorganize
the gauge group.
The appearance of mixed states is in general controlled by the
phase $\cc{z_1}{z_2}$. The choice $\cc{z_1}{z_2}=+1$ allows for
mixed gauge bosons
and leads to the gauge groups presented in Table \ref{gga}.
\begin{table}
\centering
\begin{tabular}{|c|c|c|c|c|c|c|c|c|}
\hline
$\cc{z_1}{z_2}$&$\cc{b_1}{z_1}$&$\cc{b_2}{z_1}$&$\cc{b_1}{z_2}$&$\cc{b_2}{z_2}$&$\cc{e_1}{z_1}$&$\cc{e_2}{z_2}$
&$\cc{e_1}{e_2}$&Gauge group\\
\hline
$+$&$+$&$+$&$+$&$+$&$+$&$+$&$+$&$SO(10)\times{SO(18)}\times{U(1)^2}$\\
\hline
$+$&$+$&$+$&$+$&$+$&$-$&$-$&$+$&$SO(10)\times{SO(9)}^2\times{U(1)^3}$\\
\hline
$+$&$+$&$+$&$+$&$+$&$-$&$+$&$+$&$SO(10)^2\times{SO(9)}\times{U(1)^2}$\\
\hline
$+$&$+$&$+$&$+$&$-$&$+$&$+$&$+$&$SO(10)^3\times{U(1)}$\\
\hline
$+$&$-$&$-$&$-$&$-$&$+$&$+$&$+$&$SO(26)\times{U(1)}^3$\\
\hline
$-$&$+$&$+$&$+$&$+$&$+$&$+$&$+$&$E_6\times{U(1)^2}\times{E_8}$\\
\hline
$-$&$-$&$+$&$-$&$+$&$+$&$+$&$+$&$E_6\times{U(1)^2}\times{SO(16)}$\\
\hline
$-$&$-$&$+$&$+$&$-$&$+$&$+$&$+$&$E_6\times{U(1)^2}\times{SO(8)\times{SO(8)}}$\\
\hline
$-$&$+$&$+$&$+$&$+$&$+$&$+$&$-$&$SO(10)\times{U(1)}^3\times{E_8}$\\
\hline
$-$&$+$&$+$&$+$&$+$&$-$&$-$&$-$&$SO(10)\times{U(1)}^3\times{SO(16)}$\\
\hline
\end{tabular}
\caption{\label{gga}\it Typical enhanced gauge groups  and
associated projection coefficients for a generic model generated
by the basis (\ref{basis})(coefficients not included
equal to +1 except those fixed by space-time supersymmetry
and conventions).
}
\end{table}

The choice  $\cc{z_1}{z_2}=-1$ eliminates all mixed gauge
bosons and there are a few possible enhancements:
${SO(10)}\times{U(1)}\to{E_6}$ and/or ${SO(8)}^2\to\{SO(16),E_8\}$.
The $x$ sector gauge bosons survive only when
\ba
&&\sum_{j=1,i\ne j}^6 \oo{e_i}{e_j}+\sum_{k=1}^2\oo{e_i}{z_k}=0\ \mod 2  \ , \ i=1,\dots,6\label{gb1}\\
&&
\sum_{j=1}^6\oo{e_j}{z_k}=0\ \mod 2\ , \ k=1,2\label{gb2}
\ea
where we have introduced the notation
\ba
\cc{v_i}{v_j}=e^{i \pi \oo{v_i}{v_j}}\,\ ,\  \oo{v_i}{v_j}=0,1
\ea
and one of the constraints in (\ref{gb1}),(\ref{gb2})  can be dropped because is linearly independent with the rest.

As far as the ${SO(8)}^2$ is concerned we have the following
possibilities:
\begin{eqnarray}
\text{(i)} & & \oo{e_i}{z_1}=\oo{b_a}{z_1}=0\  \forall\  i=1,\dots,6,\ a=1,2\\
\text{(ii)} & & \oo{e_i}{z_2}=\oo{b_a}{z_2}=0\  \forall\  i=1,\dots,6,\  a=1,2\\
\text{(iii)} & & \oo{e_i}{z_1+z_2}=\oo{b_a}{z_1+z_2}=0\  \forall\
i=1,\dots,6,\  a=1,2\label{ge}
\end{eqnarray}
Depending on which of the above equations are true the enhancement
is
\begin{eqnarray}
&&\mbox{both } \text{(i)} \mbox{ and } \text{(ii)}  \Longrightarrow {E_8}\\
&&\mbox{one of } \text{(i)} \mbox{ or } \text{(ii)} \mbox{ or } \text{(iii)} \Longrightarrow {SO(16)}\\
&&\mbox{none of } \text{(i)} \mbox{ or } \text{(ii)} \mbox{ or }
\text{(iii)} \Longrightarrow {SO(8)}\times{SO(8)}
\end{eqnarray}

In the sequel we will restrict in the case $\cc{z_1}{z_2}=-1$ as
this is the more promising phenomenologically, we intent to
examine $\cc{z_1}{z_2}=+1$ in detail in a future publication.


\subsection{Observable matter spectrum}
The untwisted sector matter is common to all models
and consists of six vectorials of $SO(10)$ and  12 non-Abelian gauge
group singlets.
In models where the gauge group  enhances to $E_6$ extra matter
comes from the $x$ sector
giving rise to six  $E_6$ fundamental reps ($\bf 27$).

Chiral twisted matter arise from the following  sectors
\begin{eqnarray}
B_{pqrs}^{(1)}&=&S+b_1+p\,e_3+q\,e_4 +
r\,e_5 + s\,e_6+(x) \nn\\
B_{pqrs}^{(2)}&=&S+b_2+p\,e_1+q\,e_2 +
r\,e_5 + s\,e_6+(x) \label{ss}\\
B_{pqrs}^{(3)}&=& S + b_3+ p\,e_1+q\, e_2 +r\,e_3+ s\,e_4+(x)\nn
\end{eqnarray}
where $b_3=b_1+b_2+x$. These are 48 sectors (16 sectors per orbifold
plane) and we choose to label them
using  the plane number $i$ (upper index) and  the integers
$p_i,q_i,r_i,s_i=\{0,1\}$ (lower index) corresponding to the
coefficients of the appropriate shift vectors. Note that for a particular
orbifold plane $i$ only
four shift vectors can be added to the twist vector $b_i$ (the ones that
have non empty intersection)
the other two give rise to massive states. Each of the above sectors
(\ref{ss}) can produce
a single spinorial of $SO(10)$ (or fundamental of $E_6$ in the case of
enhancement). Since the $E_6$ model
spectrum is in one to one correspondence with the $SO(10)$ spectrum in
the following we use the
name spinorial meaning the $\bf 16$ of $SO(10)$ and in the case of
enhancement the $\bf27$ of $E_6$.

One of the  advantages of our formulation is that it allows to extract
generic  formulas regarding the
number and the chirality of each spinorial. This is important because
it allow a algebraic
treatment of the entire class of models without deriving each model
explicitly. The number of
surviving spinorials per sector (\ref{ss}) is given by
\begin{eqnarray}
P_{pqrs}^{(1)}&=&
\frac{1}{16}\,\prod_{i=1,2}\left(1-\cc{e_i}{B_{pqrs}^{(1)}}\right)\,
\prod_{m=1,2}\left(1-\cc{z_m}{B_{pqrs}^{(1)}}\right)\,\label{pa}\\
P_{pqrs}^{(2)}&=&
\frac{1}{16}\,\prod_{i=3,4}\left(1-\cc{e_i}{B_{pqrs}^{(2)}}\right)\,
\prod_{m=1,2}\left(1-\cc{z_m}{B_{pqrs}^{(2)}}\right)\,\label{pb}
\\
P_{pqrs}^{(3)}&=&
\frac{1}{16}\,\prod_{i=5,6}\left(1-\cc{e_i}{B_{pqrs}^{(3)}}\right)\,
\prod_{m=1,2}\,\left(1-\cc{z_m}{B_{pqrs}^{(3)}}\right)\,\label{pc}
\end{eqnarray}
where $P_{pqrs}^i$ is a projector that takes values $\{0,1\}$.
The chirality of the surviving spinorials is given by
\begin{equation}
X_{pqrs}^{(1)}=\cc{b_2+(1-r) e_5+(1-s) e_6}{B_{pqrs}^{(1)}}\label{ca}
\end{equation}
\begin{equation}
X_{pqrs}^{(2)}=\cc{b_1+(1-r) e_5+(1-s) e_6}{B_{pqrs}^{(2)}}\label{cb}
\end{equation}
\begin{equation}
X_{pqrs}^{(3)}=\cc{b_1+(1-r) e_3+(1-s) e_4}{B_{pqrs}^{(3)}}\label{cc}
\end{equation}
where $X_{pqrs}^i=+$ corresponds to a ${\bf16}$ of $SO(10)$(or ${\bf27}$
in the case of $E_6$) and $X_{pqrs}^i=-$ corresponds
to a ${\bf\overline{16}}$ (or ${\bf\overline{27}}$) and we have chosen
the space-time chirality $C(\psi^\mu)=+1$.
The net number of spinorials and thus the net number of families is given by
\begin{equation}
N_F=\sum_{i=1}^3\sum_{p,q,r,s=0}^1X^{(i)}_{pqrs}
P_{pqrs}^{(i)}\label{nf}
\end{equation}
Similar formulas can be easily derived for the number of vectorials
and the number of singlets and can be extended
to the $U(1)$ charges but in this work we will restrict to the
spinorial calculation.

Formulas (\ref{pa})-(\ref{pc}) allow us to identify the mechanism
of spinorial reduction, or in other words
the fixed point reduction, in the fermionic language. For a particular
sector ($B_{pqrs}^{(i)}$) of the orbifold plane $i$
there exist two shift vectors ($e_{2i-1}, e_{2i}$)
and the two zeta vectors ($z_1,z_2$) that have no common elements
with $B_{pqrs}^{(i)}$. Setting the relative projection coefficients (\ref{pc})
to $-1$ each of the above four vectors acts as a projector that cuts
the number of fixed points in the
associated sector by a factor of two. Since four such projectors are
available for each sector the number of fixed points
can be reduced from $16$ to $1$ per plane.

The projector action (\ref{pa})-(\ref{pc}) can be expanded and
written in a simpler form
\ba
\Delta^{(i)} W^{(i)}=Y^{(i)} \label{proj}
\ea
where
\ba
\Delta^{(1)}&=&\left[
\begin{array}{cccc}
\oo{e_1}{e_3}&\oo{e_1}{e_4}&\oo{e_1}{e_5}&\oo{e_1}{e_6}\\
\oo{e_2}{e_3}&\oo{e_2}{e_4}&\oo{e_2}{e_5}&\oo{e_2}{e_6}\\
\oo{z_1}{e_3}&\oo{z_1}{e_4}&\oo{z_1}{e_5}&\oo{z_1}{e_6}\\
\oo{z_2}{e_3}&\oo{z_2}{e_4}&\oo{z_2}{e_5}&\oo{z_2}{e_6}
\end{array}
\right]
\ ,\
Y^{(1)}=
\left[
\begin{array}{c}
\oo{e_1}{b_1}\\
\oo{e_2}{b_1}\\
\oo{z_1}{b_1}\\
\oo{z_2}{b_1}
\end{array}
\right]
\nn
\\
\Delta^{(2)}&=&\left[
\begin{array}{cccc}
\oo{e_3}{e_1}&\oo{e_3}{e_2}&\oo{e_3}{e_5}&\oo{e_3}{e_6}\\
\oo{e_4}{e_1}&\oo{e_4}{e_2}&\oo{e_4}{e_5}&\oo{e_4}{e_6}\\
\oo{z_1}{e_1}&\oo{z_1}{e_2}&\oo{z_1}{e_5}&\oo{z_1}{e_6}\\
\oo{z_2}{e_1}&\oo{z_2}{e_2}&\oo{z_2}{e_5}&\oo{z_2}{e_6}
\end{array}
\right]
\ ,\
Y^{(2)}=
\left[
\begin{array}{c}
\oo{e_3}{b_2}\\
\oo{e_4}{b_2}\\
\oo{z_1}{b_2}\\
\oo{z_2}{b_2}
\end{array}
\right]
\\
\Delta^{(3)}&=&\left[
\begin{array}{cccc}
\oo{e_5}{e_1}&\oo{e_5}{e_2}&\oo{e_5}{e_3}&\oo{e_5}{e_4}\\
\oo{e_6}{e_1}&\oo{e_6}{e_2}&\oo{e_6}{e_3}&\oo{e_6}{e_4}\\
\oo{z_1}{e_1}&\oo{z_1}{e_2}&\oo{z_1}{e_3}&\oo{z_1}{e_4}\\
\oo{z_2}{e_1}&\oo{z_2}{e_2}&\oo{z_2}{e_3}&\oo{z_2}{e_4}
\end{array}
\right]
\ ,\
Y^{(3)}=
\left[
\begin{array}{c}
\oo{e_5}{b_3}\\
\oo{e_6}{b_3}\\
\oo{z_1}{b_3}\\
\oo{z_2}{b_3}
\end{array}
\right]
\nn
\ea
and
\ba W^i= \left[
\begin{array}{c}
p_i\\
q_i\\
r_i\\
s_i
\end{array}
\right]
\ea
They form three systems of equations of the form $\Delta^i\,W^i=Y^i$
(one for each orbifolds plane).
Each system contains  4 unknowns $p_i,q_i,r_i,s_i$ which correspond
to the
labels of surviving spinorials in the plane $i$. We call the set of
solutions of each system
$\Xi_i$. The net number of families (\ref{nf}) can be written as
\ba
N_F=\sum_{i=1}^3\sum_{p,q,r,s\in\Xi_i}X^{(i)}_{pqrs}
\ea
The chiralities (\ref{ca})-(\ref{cc}) can be further expanded in the
exponential form
$X_{pqrs}^{(i)}=\exp \left( i \pi \chi_{pqrs}^{(i)}\right)$
\ba
\chi_{pqrs}^{(1)} &=& 1+\oo{b_1}{b_2} + (1-r)\oo{e_5}{b_1}
+ (1-s) \oo{e_6}{b_1} + p \oo{e_3}{b_2}
\nn\\
&&+ q \oo{e_4}{b_2}
+ r \oo{e_5}{b_2} + s \oo{e_6}{b_2}
+ p (1-r) \oo{e_3}{e_5}
\nn\\
&& + p (1-s) \oo{e_3}{e_6}
+ q (1-r) \oo{e_4}{e_5}
+ q (1-s) \oo{e_4}{e_6}\nn\\
&&+ (r+s) \oo{e_5}{e_6} \mod 2
\label{chia}
\ea
\ba
\chi_{pqrs}^{(2)} &=& 1+\oo{b_1}{b_2} + (1-r)\oo{e_5}{b_2}
+ (1-s) \oo{e_6}{b_2} + p \oo{e_1}{b_1}
\nn\\
&&+ q \oo{e_2}{b_1}
+ r \oo{e_5}{b_1} + s \oo{e_6}{b_1}
+ p (1-r_2) \oo{e_1}{e_5}
\nn\\
&& + q (1-r) \oo{e_2}{e_5}
+ p (1-s) \oo{e_1}{e_6}
+ q (1-s) \oo{e_2}{e_6}\nn\\
&&+ (r+s) \oo{e_5}{e_6} \mod 2
\label{chib}
\ea
\ba
\chi_{pqrs}^{(3)} &=& 1+\oo{b_1}{b_2} + (1-p)\oo{e_1}{b_1}
+ (1-q) \oo{e_2}{b_1} + \oo{e_5+e_6}{b_1}
\nn\\
&& + (1-r) \oo{e_3}{b_2} + (1-s) \oo{e_4}{b_2}\nn\\
&& + (1-r)(1-p) \oo{e_3}{e_1} + (1-r)(1-q_3) \oo{e_3}{e_2}\nn\\
&&+ (1-r) \oo{e_3}{e_5} + (1-r) \oo{e_3}{e_6} + (1-s) \oo{e_4}{e_6}
\nn\\
&&+ (1-r) \oo{e_3}{z_1+z_2} + (1-s) \oo{e_4}{z_1+z_2}\nn\\
&&+\oo{b_1}{z_1+z_2} \mod 2
\label{chic}
\ea
We remark here that the projection coefficient $\cc{b_1}{b_2}$ simply
fixes the overall chirality
and that our equations depend only on
\ba
\oo{e_i}{e_j},\ \oo{e_i}{b_A},
\oo{e_i}{z_n},\ \oo{z_n}{b_A},
i=1,\dots,6, A=1,2, n=1,2.\label{allp}
\ea
However, the following six parameters do not appear in the expressions
$\oo{e_1}{e_2},\oo{e_3}{e_4},\oo{e_3}{b_1},\oo{e_4}{b_1},\oo{e_1}{b_2},\oo{e_2}{b_2}$
and thus a generic model depends on 37 discrete parameters.

\section{Results}\label{five}
\subsection{Models \label{models}}
We apply here the formalism developed above in order to derive sample models in the free
fermionic formulation.
\bigskip

\noindent{\it The $Z_2\times Z_2$ symmetric orbifold}\\[5pt]
The simplest example is the symmetric $Z_2\times{Z_2}$ orbifold.  Here we set
all the free  GSO  phases (\ref{allp})  to zero. The full GSO phase matrix takes
the form ($\cc{v_i}{v_j}=\exp[i\pi(v_i|v_j)]$)
{\small$$
(v_i|v_j)\ \ =\ \
\bordermatrix{
&1 & S & e_1 & e_2 & e_3 & e_4 & e_5 & e_6 & b_1 & b_2 & z_1 & z_2\cr
   1   & 1 & 1  &  1 &  1 &  1 &  1 &  1 &  1 &  1 &  1 &  1 & 1 \cr
   S   & 1 & 1  &  1 &  1 &  1 &  1 &  1 &  1 &  1 &  1 &  1 & 1\cr
  e_1  & 1 &  1 &  0 &  0 &  0 &  0 &  0 &  0 &  0 &  0 &  0 & 0\cr
  e_2  & 1 &  1 &  0 &  0 &  0 &  0 &  0 &  0 &  0 &  0 &  0 & 0\cr
  e_3  & 1 &  1 &  0 &  0 &  0 &  0 &  0 &  0 &  0 &  0 &  0 & 0\cr
  e_4  & 1 &  1 &  0 &  0 &  0 &  0 &  0 &  0 &  0 &  0 &  0 & 0\cr
  e_5  & 1 &  1 &  0 &  0 &  0 &  0 &  0 &  0 &  0 &  0 &  0 & 0\cr
  e_6  & 1 &  1 &  0 &  0 &  0 &  0 &  0 &  0 &  0 &  0 &  0 & 0\cr
  b_1  & 1 &  0 &  0 &  0 &  0 &  0 &  0 &  0 &  1 &  1 &  0 & 0\cr
  b_2  & 1 &  0 &  0 &  0 &  0 &  0 &  0 &  0 &  1 &  1 &  0 & 0\cr
  z_1  & 1 &  1 &  0 &  0 &  0 &  0 &  0 &  0 &  0 &  0 &  1 & 1\cr
  z_2  & 1 &  1 &  0 &  0 &  0 &  0 &  0 &  0 &  0 &  0 &  1 & 1\cr
  }
$$}
With the above choice $\Delta^{(i)}=W^{(i)}=0$ in  equation (\ref{proj}).
All projectors become inactive and thus the number of surviving twisted
sector spinorials takes its maximum
value which is 48 with all chiralities positive according to
(\ref{chia}), (\ref{chib}), (\ref{chic}).
 Moreover three spinorials and three anti-spinorials arise from the
untwisted sector.
Following (\ref{gb1}), (\ref{gb2}) the gauge group enhances to
$E_6\times{U(1)}^2\times{E_8}$ and the spinorials of $SO(10)$
combine with vectorials and singlets to produce 48+3=51 families ($\bf 27$) and 3 anti-families ($\bf \overline{27}$) of $E_6$.
\bigskip

\noindent{\it A three generation $E_6$ model}\\[5pt]
We can obtain a three family $E_6$ model by choosing the following set of projection coefficients
{\small$$
(v_i|v_j)\ \ =\ \
\bordermatrix{
       &1  &S   & e_1& e_2& e_3& e_4&e_5 & e_6& b_1& b_2&z_1 &z_2\cr
   1   &1  &1  &1  &1  &1  &1  &1  &    1  &    1  &    1  &    1  &  1\cr
   S   &1  &1  &1  &1  &1  &1  &1  &    1  &    1  &    1  &    1  &  1\cr
  e_1  &1  &1  &0  &0  &1  &0  &0  &    1  &    0  &    0  &    0  &  0\cr
  e_2  &1  &1  &0  &0  &0  &0  &1  &    0  &    0  &    0  &    0  &  1\cr
  e_3  &1  &1  &1  &0  &0  &0  &1  &    0  &    0  &    0  &    0  &  0\cr
  e_4  &1  &1  &0  &0  &0  &0  &1  &    0  &    0  &    0  &    1  &  0\cr
  e_5  &1  &1  &0  &1  &1  &1  &0  &    1  &    0  &    0  &    0  &  0\cr
  e_6  &1  &1  &1  &0  &0  &0  &1  &    0  &    0  &    0  &    1  &  1\cr
  b_1  &1  &0  &0  &0  &0  &0  &0  &    0  &    1  &    0  &    0  &  0\cr
  b_2  &1  &0  &0  &0  &0  &0  &0  &    0  &    0  &    1  &    0  &  0\cr
  z_1  &1  &1  &0  &0  &0  &1  &0  &    1  &    0  &    0  &    1  &  1\cr
  z_2  &1  &1  &0  &1  &0  &0  &0  &    1  &    0  &    0  &    1  &  1\cr
  }
$$}
The full gauge group is here $E_6\times {U(1)}^2\times {SO(8)}^2$. Three families
$({\bf 27})$, one
from each plane,  arise from the sectors $S+b_i+(x), i=1,2,3$. Another set of
three families and three
anti-families arise from the untwisted sector.
The hidden sector consists of  nine 8-plets under each $SO(8)$. In addition
there exist a number of non-Abelian
gauge group singlets. The model could be phenomenologically acceptable
provided one finds a way
of breaking $E_6$. Since it is not possible to generate the $E_6$ adjoint
(not in Kac-Moody level one),
we need to realize the breaking by an additional Wilson-line like vector.
However,  a detailed
investigation of acceptable basis vectors, shows that the $E_6$ breaking
is accompanied by
truncation of the fermion families. Thus this kind of perturbative $E_6$
breaking is not compatible
with the presence of three generations. It would be interesting to utilize
string dualities in order to study
the non-perturbative aspects of such models.
\bigskip

\noindent{\it A six generation $E_6$ model}\\[5pt]
Similarly a six family $E_6\times{U(1)}^2\times{E_8}$ model can be obtained using the following
projection coefficients
{\small$$
(v_i|v_j)\ \ =\ \
\bordermatrix{
      &1  &S   & e_1& e_2& e_3& e_4&e_5 & e_6& b_1& b_2&z_1 &z_2\cr
   1  &1  &   1  &   1  &   1  &   1  &   1  &   1  &   1  &   1  &   1  &   1&   1\cr
   S  &1  &   1  &   1  &   1  &   1  &   1  &   1  &   1  &   1  &   1  &   1&   1\cr
  e_1 &1  &   1  &   0  &   0  &   0  &   0  &   1  &   1  &   0  &   0  &   0&   0\cr
  e_2 &1  &   1  &   0  &   0  &   1  &   0  &   0  &   1  &   0  &   0  &   0&   0\cr
  e_3 &1  &   1  &   0  &   1  &   0  &   0  &   0  &   1  &   0  &   0  &   0&   0\cr
  e_4 &1  &   1  &   0  &   0  &   0  &   0  &   1  &   0  &   0  &   0  &   0&   1\cr
  e_5 &1  &   1  &   1  &   0  &   0  &   1  &   0  &   0  &   0  &   0  &   0&   0\cr
  e_6 &1  &   1  &   1  &   1  &   1  &   0  &   0  &   0  &   0  &   0  &   0&   1\cr
  b_1 &1  &   0  &   0  &   0  &   0  &   0  &   0  &   0  &   1  &   0  &   0&   0\cr
  b_2 &1  &   0  &   0  &   0  &   0  &   0  &   0  &   0  &   0  &   1  &   0&   0\cr
  z_1 &1  &   1  &   0  &   0  &   0  &   0  &   0  &   0  &   0  &   0  &   1&   1\cr
  z_2 &1  &   1  &   0  &   0  &   0  &   1  &   0  &   1  &   0  &   0  &   1&   1\cr
 }
$$}
In this model we have six families from the twisted sector, two from each plane together with
three families and three anti-families from the untwisted sector, accompanied by a number of singlets and
8-plets of both hidden $SO(8)$'s.

\subsection{$N=4$ lift-able vacua}
In the models considered above we have managed to separate the  orbifold
twist action (represented here by $b_1$, $b_2$) from
the shifts (represented by $e_i$) and the Wilson lines $(z_1,z_2)$. However,
these actions are further correlated
through the GSO projection coefficients $\cc{v_i}{v_j}$.  Nevertheless, we
remark that the twist action can be
 decoupled from the other two in the case
\begin{equation}
\cc{b_n}{z_k}=\cc{b_m}{e_i}=+1\ , i=1,\dots,6,\ k=1,2,\ m,n=1,2,3\label{lifte}
\end{equation}
The above relation defines a subclass of $N=1$ four dimensional vacua with
interesting phenomenological properties
and includes three generation models. Due to the decoupling of the orbifold
twist action these vacua are direct descendants
of $N=4$ vacua so we will refer to these models as $N=4$ lift-able models.
In this subclass of models  some important phenomenological properties of
the vacuum, as the number of generations, are
predetermined at the $N=4$ level as it is related to the
$\oo{e_i}{e_j}$ and $\oo{z_i}{e_j}$ phases.
The orbifold action reduces the supersymmetries and the gauge group and
makes chirality apparent, however the number of generations is selected by
the $N=4$ vacuum structure. At the
$N=1$ level this is understood as follows: the $Z_2\times{Z_2}$ orbifold
has $48$ fix points.
Switching on some of the above phase correspond to a free action that
removes some of the fixed points
and thus reduces the number of spinorials. Moreover, in this case,
the chirality of the surviving spinorials
is again related as seen by (\ref{chia})-({\ref{chic}) to the $\oo{e_i}{e_j}$
and $\oo{e_i}{z_k}$ coefficients, which are
all fixed at the $N=4$ level.
The observable gauge group of liftable models is always $E_6$ and this can
be easily seen by applying (\ref{lifte}) to (\ref{gb1}),(\ref{gb2}).

Typical examples of   such vacua are the three and six generation
$E_6\times {U(1)}^2\times {SO(8)}^2$ models presented
in section \ref{models}.
A careful counting, taking into account some symmetries among the
coefficients, shows that this class of models consists of $2^{20}$
models, or $2^{21}$ if we include $\oo{b_1}{b_2}$.
These vacua are interesting because they can admit a geometrical interpretation.

From the orbifold description we learn that all breakings of the
hidden and observable gauge group are induced using Wilson lines. From the
$4D$  point of view the internal gauge group is broken in a similar fashion
using Wilson lines. The twisted planes in equation \eqref{eq:twisted}
describe the
removal of the free moduli using twists. When a group is broken using Wilson
lines the field corresponding to this Wilson line obtains a nonzero VEV.
The fixing of
the moduli using twists can be interpreted as the removal of the quantum
fluctuations of the fields identified with the Wilson lines.
These Wilson lines
become discrete Wilson lines and the VEV becomes a fixed value.

\subsection{Classification}

As we discussed above, the free GSO phases of the $N=1$ partition
function control the number of chiral generations in a given
model. In section 3 we have given analytic formulas that enable
the calculation of the number of generations for any given set of
phases. To gain an insight to the structure of this class of vacua
we can proceed with a computer evaluation of these formulas and
thus classify the space of these vacua with respect to the number
of generations. This also allows detailed examination of the
structure of these vacua and in particular how the generations are
distributed among the three orbifold planes. The main obstacle to
this approach is the huge number of vacua under consideration. As
a first step in this direction we restrict ourselves to the class
of lift-able  vacua that is  physically appealing and contains
representative models with the right number of generations. As
stated above this class consists in principle of $2^{21}$
models
and their complete classification  takes a few minutes on a personal computer
using an appropriate computer program.
The  program analyses all different options for the free GSO coefficients.
The different configurations are then used to calculate the number of
generations
using formulae \eqref{pa} -- \eqref{nf}. For the analysis
of the gauge group we use formulae \eqref{gb1} -- \eqref{ge}.
The results are presented in Tables \ref{tab:real11} -- \ref{tab:real3}.
In these tables we list the number of generations coming from the twisted
sectors. They are listed per plane. The number of positive chiral
generations is separated from the number of negative chiral generations
on each plane. The total number is then listed before listing the total
net number of generations. As the sign of the chirality is determined by
the coefficient $\oo{b_1}{b_2}$ (see \eqref{chia} -- \eqref{chic}) we
have included models that have a positive net number of generations.
In order to maintain a complete separation of the hidden gauge group we
have set $\oo{z_1}{z_2}=1$. The tables are ordered by the total net number
of chiral states.

We find that there are no liftable models with a $SO(10)$
observable gauge group, which is always extended to $E_6$,
and the states from the vector $x$ are not projected out.
Since the models admit a geometrical interpretation, it means that they
must descend from the ten dimensional  $E_8\times E_8$
heterotic--string on $Z_2\times Z_2$ Calabi--Yau threefold.

In $3\%$ of all the models the hidden gauge group
is enhanced to $SO(8)\times SO(8) \to SO(16)$. We find that in total $1024$
liftable models are enhanced to $SO(8)\times SO(8) \to E_8$. We find that
the 24 generations NAHE model as explained in section \ref{two} is present
in Table \ref{tab:real11}.
The problem of a detailed investigation of the full class of
vacua will be considered further in a future publication.

\subsection{General properties}
In section \ref{three} we discussed a direct translation between the
bosonic
formulation and the fermionic formulation of the heterotic string
compactifications. $Z_2\times Z_2$ orbifold
compactifications are relevant for our class of models.
These orbifolds contain three twisted
sectors, or three twisted planes. A priori we may have the
possibility that all three twisted planes produce spinorial
$SO(10)$ representations. We refer to this sub--class of models
as $S^3$ models. The alternatives are models in which spinorial
representations may be obtained from only two, one, or none, twisted
planes, and the others produce vectorial representations. We refer
to these cases as $S^2V$, $SV^2$ and $V^3$ models, respectively.
The focus of the analysis in this paper is on the $S^3$
sub--class of models, which also contains the NAHE--based
three generation models. The $S^3$ sub--class allows, depending
on the one--loop GSO projection coefficients, the possibility
of spinorials on each plane. In specific models in this sub--class
each Standard Model family is obtained from a distinct orbifold
plane. Such models therefore produce three generation models
and may be phenomenologically interesting.
The only other phenomenologically viable option can come from the subclass
$S^2V$ models as this class of models may contain a model with for example $2$
generations coming from the first plane and $1$ generation coming form the
second plane and none from the third. The $SV^2$ class of models cannot
produce a
physical model because it is not possible to reduce the number of families to
$3$ as they would have to be coming from one plane and $3$ cannot be written
as a power of $2$. Similarly the $V^3$ subclass of models will not contain
phenomenologically interesting models.

\paragraph{$3$ generations realized only with twisted and shifted real
manifolds.}
Since the projectors are constructed using the complete separation of the
internal manifold we see that three generation models are only possible when
\begin{equation}
\Gamma_{6,6}\  =\  \Gamma_{2,2}^3\ \  \longrightarrow\ \  \Gamma_{1,1}^6.
\end{equation}
These $\Gamma_{1,1}$ internal parts do not describe a complex manifold. They
describe internal real circles. If we use solely complex manifolds,
of the type $\Gamma_{6,6}=\Gamma_{2,2}^3$, and using only
symmetric shifts, we find that
there are no $3$ generation models.
We therefore conclude that the net number of generations can
never be equal to three
in the framework of $Z_2\times Z_2$ Calabi--Yau compactification.
This implies the necessity of non zero torsion in CY $Z_2\times Z_2$
compactifications in order to obtain semi--realistic three
generation models.

In the realistic free fermionic
models the reduction of the number of families together with
the breaking of the observable $SO(10)$ is realized by isolating full
multiplets at two fixed points on the internal manifold.
In reducing the number of families
down to one, different component of each family are attached
to the two distinct fixed points.
We remove one full multiplet and simultaneously break the
$SO(10)$ symmetry. We
therefore keep a full multiplet on each twisted plane.
In the $SO(10)$ models described here a whole $16$ or $\overline{16}$
of $SO(10)$ is attached to a fixed point.
We are therefore not able to break the $SO(10)$, and simultaneously
preserve the full Standard Model multiplets.
For this reason we find that the
observable $SO(10)$ cannot be broken perturbatively
in this class of three generation models, and may only
be broken nonperturbatively.
It is therefore not possible to reduce both the number of
families down to $3$ and break the observable gauge group $SO(10)$ down
to its subgroups perturbatively.

We conclude that there is a method to reduce the number of generations from
$48$ to $3$. Since we need $4$ projectors we need to separate the hidden gauge
group using $SO(8)$ characters
\begin{equation}
\Gamma_{0,8} \to \Gamma_{0,4}\ \Gamma_{0,4}
\end{equation}
and we need to break the internal complex manifold to an internal
real manifold
\begin{equation}
\Gamma_{6,6} \to \left[ \Gamma_{1,1}\ \Gamma_{1,1} \right]^3.
\end{equation}
If we reduce the number of generations to $3$ we cannot break the $SO(10)$
observable group to its subgroups, while maintaining a full multiplet. The
$SO(10)$
observable gauge group cannot therefore be broken perturbatively.
We can reduce the number of generations from $48$ to $6$ using $3$ projectors.
This entails that we can choose either to separate the hidden gauge group
using $SO(16)$ characters, or to leave the internal manifold complex.


We argued above that we cannot break $SO(10)$ down to
a subgroup perturbatively, while reducing the number of generations
to $3$.
If we want to break the $SO(10)$ symmetry perturbatively,
and keep a full $SO(10)$ multiplet from a given twisted plane,
we can only reduce the number of generations to $6$.
This can be achieved if we define three different
projectors like the ones defined in equations \eqref{pa} -- \eqref{pc}.
We are therefore left with two options.
\begin{itemize}
\item We can use $SO(16)$ characters for the separation of the hidden gauge
group. We have then constructed only one projector which leaves us no other
 option
 than to break the complex structure using symmetric shifts
\begin{equation}
\Gamma_{6,6} \to \Gamma_{1,1}^6.
\end{equation}
\item We can use $SO(8)$ characters for the separation of the hidden gauge
group. In doing so we have constructed two projectors. The third can be
realized
by the symmetric shifts that leave the complex structure of the internal
manifold
 intact
\begin{equation}
\Gamma_{6,6} \to \Gamma_{2,2}^3.
\end{equation}
\end{itemize}
\vspace{\baselineskip}

\section{Discussion and conclusions}\label{conclusion}

String theory duly attracts wide interest. It provides a
consistent approach to perturbative quantum gravity,
while at the same time incorporating the gauge and
matter structures that are relevant for particle physics
phenomenology. However, the multitude of vacua that the
theory admits and the lack of a dynamical principle to choose
among them, hinders the prospects that the theory will yield
unique experimental predictions. This has led some authors
to advocate the anthropic principle as a possible resolution for
understanding the contrived set of parameters that seem to
govern our world.

The approach pursued in our work is different. In our view
the understanding of the dynamical principles that underly
quantum gravity and the vacuum selection must await the
better conceptual understanding of the quantum gravity synthesis.
It may well be that at the end of the day the probabilistic
nature of quantum mechanics will emerge as a derived
property rather than a fundamental property of quantum gravity.
In this respect we should regard the string theories as
merely providing a perturbative glimpse into the underlying
properties of quantum gravity, and how it may relate to the
gauge and matter observables. In this context we must utilize
both the low energy data as well as the basic properties
of string theory to isolate promising string vacua and
develop the tools to discern between the experimental
predictions of different classes. An example,
is the $SO(10)$ embedding of the Standard Model
spectrum, which is viable in the heterotic limit of
M--theory, but not in its type I limit.

Given the Standard Model properties, we may hypothesize that the
true string vacuum should accommodate two pivotal ingredients.
One is the existence of three generations and the second is
their embedding in an
underlying $SO(10)$ or $E_6$ grand unified gauge group.
In this context, the replication of the matter generations
is the first particle physics observable whose origin may be
sought in string theory. This follows from the fact that the
flavor sector of the Standard Model does not arise from
any physical principle, like the gauge principle, as well
as from the fact that in certain classes of string compactifications
the number of generations is related to a topological number
of compact manifolds, the Euler characteristic.

A class of string compactifications that admit three generations,
as well as their embedding in an underlying $SO(10)$ group structure
are the NAHE--based free fermionic models. A subset of the boundary
condition basis vectors that span these models can be seen to correspond
to $Z_2\times Z_2$ orbifold compactifications at special points in the
moduli space. However, the geometrical understanding of the full three
generation  models is still lacking. The aim of the current work
is therefore to advance the geometrical understanding of the
NAHE--based free fermionic models.
In this paper we showed that two
of the boundary condition basis vectors beyond those that correspond
to the $Z_2\times Z_2$ orbifold correspond to symmetric shifts
on the compact tori, whereas the third correspond to an asymmetric
shift. We then proceeded to classify all possible symmetric shifts
on complex tori and demonstrated that three generation models do not
arise in this manner. Three generation models that realise the $Z_2\times
Z_2$ orbifold picture of the three chiral generations were found.
In these cases the $SO(10)$ gauge group cannot be broken perturbatively,
while preserving the full Standard Model matter content. Additionally in
this cases the internal lattice is broken to $\Gamma_{1,1}^6$,
{\it i.e.} to a product of six circles. In this class of models
each of the chiral generations is attached to a single fixed point
in each of the twisted orbifold planes. This should be contrasted
with the case of the three generation free fermionic models in
which the $SO(10)$ symmetry is broken perturbatively by Wilson lines.
In those cases, each generation is obtained from a separate orbifold
plane, but different components of each generation are attached to
different fixed points of the corresponding twisted sector.

Additionally, we demonstrated in this paper that for a wide range
of models, for which a geometrical origin is understood, there exist
an interpretation of the phases that appear in the $N=1$ partition
function, in terms of vacuum expectation values of background fields
of the $N=4$ vacua. In these cases the dynamical components
of the background fields are projected out, but their vacuum expectation
value is retained and takes the form of the free GSO phases of the
$N=1$ partition function. These phases also control the chirality of the
models. Thus, we have the situation in which the chirality of the
models is already determined by the VEVs of the background fields
of the $N=4$ vacuum. In effect, the chiral content of the $N=1$ vacua
in these cases is determined by the Narain $N=4$ lattice.
An example of this phenomenon was already seen in the case of
$Z_2\times Z_2$ on $SO(12)$ lattice that yields 24 generations
versus the $Z_2\times Z_2$ orbifold on $SO(4)^3$ lattice that
yields 48 generations. The interpretation of the chiral
content of the $N=1$ models in terms of the $N=4$ vacua
will be especially instrumental when seeking the strong
coupling duals of the $N=1$ models, due to the fact that
the $N=4$ duals can be obtained with relative ease. The understanding
of the $N=1$ duals will then entail the understanding of the
corresponding $Z_2\times Z_2$ operation on the dual side.

We discovered in this paper that the three generation free
fermionic models necessarily employ an asymmetric shift
on the internal compactified space. This observation has
profound implications. In the first place, since the asymmetric
shift can act only at enhanced symmetry points in the moduli space,
it implies that some moduli are fixed and frozen. In fact in some
cases it is seen that all the geometrical moduli are projected
out. In those cases the geometrical moduli may be interchanged with
twisted moduli which are much more difficult to identify,
and hence their moduli spaces are more intricate. Additionally,
the necessity of incorporating an asymmetric shift has important
implications in the context of nonperturbative dualities.
In the case of the duals of the heterotic models,
a geometric moduli is interchanged with
the dilaton. Hence, the fact that the geometric moduli are fixed
around their self--dual value on the heterotic side implies that on the dual
side the dilaton has to be fixed around its self--dual point. This is a
fascinating possibility that we will return to in future work.
However, we note that the low energy phenomenological data may
point in the direction of esoteric compactifications
that would have otherwise been overlooked.
The results show that, in the framework of
$Z_2\times Z_2$ Calabi--Yau compactification, the net number of
generation can never be equal to
three. This implies the necessity of non zero torsion in CY
compactifications in order to obtain semi-realistic three
generation models.
Additionally, the necessity to incorporate an asymmetric shift
in the reduction to three generations, has profound implications for the
issues of moduli stabilization and vacuum selection. The reason
being that it can only be implemented at enhanced symmetry
points in the moduli space. In this context we envision that the
self--dual point under T--duality plays a special role. In the
context of nonperturbative dualities the dilaton and
moduli are interchanged, with potentially important
implications for the problem of dilaton stabilization.
We will report on these aspects in future publications.


\section*{Acknowledgements}
This  work was supported in part by the Pieter Langerhuizen
Lambertuszoon Fund of the Royal Holland Society of Sciences and Humanities,
the Noorthey Academy and the VSB foundation(SN); by the  European Union
under the
contracts HPRN-CT-2000-00122, HPRN-CT-2000-00131,
HPRN-CT-2000-00148, HPRN-CT-2000-00152 and HPMF-CT-2002-01898;
by the PPARC and by the Royal Society.




\bibliographystyle{unsrt}



\appendix

\begin{table}
\begin{center}
\begin{tabular}{r || r r | r r | r r | r r | r }
&\multicolumn{2}{c|}{$1$}&\multicolumn{2}{c|}{$2$}&\multicolumn{2}{c|}{$3$}&\multicolumn{2}{c|}{total}&\multicolumn{1}{c}{net}\\
No.  & $+$& $-$& $+$& $-$& $+$& $-$&  $+$& $-$ &  \\\hline\hline
1&16&0&8&0&8&0&32&0&32\\
2&8&0&8&0&8&0&24&0&24\\
3&8&0&8&0&4&0&20&0&20\\
4&8&0&6&2&4&0&18&2&16\\
5&8&0&4&0&4&0&16&0&16\\
6&12&4&4&0&4&0&20&4&16\\
7&8&0&8&0&4&4&20&4&16\\
8&6&2&4&0&4&0&14&2&12\\
9&4&0&4&0&4&0&12&0&12\\
10&8&0&2&0&2&0&12&0&12\\
11&4&0&4&0&2&0&10&0&10\\
12&4&0&4&0&3&1&11&1&10\\
13&6&2&4&0&2&0&12&2&10\\
14&4&4&4&0&4&0&12&4&8\\
15&4&0&4&0&2&2&10&2&8\\
16&4&0&3&1&2&0&9&1&8\\
17&4&0&2&0&2&0&8&0&8\\
18&6&2&3&1&2&0&11&3&8\\
19&6&2&2&0&2&0&10&2&8\\
20&10&6&2&0&2&0&14&6&8\\
21&6&2&4&0&2&2&12&4&8\\
22&3&1&3&1&2&0&8&2&6\\
23&3&1&2&0&2&0&7&1&6\\
24&2&0&2&0&2&0&6&0&6\\
25&4&0&2&2&2&0&8&2&6\\
26&4&0&2&0&1&1&7&1&6\\
27&4&0&1&0&1&0&6&0&6\\
28&6&2&1&0&1&0&8&2&6\\
29&3&1&3&1&1&0&7&2&5\\
30&2&0&2&0&1&0&5&0&5
\end{tabular}
\caption{\cption{$E_6 \times U(1)^2 \times SO(8) \times SO(8)$}}
\label{tab:real11}
\end{center}
\end{table}


\begin{table}
\begin{center}
\begin{tabular}{r || r r | r r | r r | r r | r }
&\multicolumn{2}{c|}{$1$}&\multicolumn{2}{c|}{$2$}&\multicolumn{2}{c|}{$3$}&\multicolumn{2}{c|}{total}&\multicolumn{1}{c}{net}\\
No.  & $+$& $-$& $+$& $-$& $+$& $-$&  $+$& $-$ &  \\\hline\hline
31&3&1&2&0&1&0&6&1&5\\
32&3&1&2&0&2&2&7&3&4\\
33&2&2&2&0&2&0&6&2&4\\
34&4&4&2&0&2&0&8&4&4\\
35&4&0&2&2&2&2&8&4&4\\
36&3&1&2&0&1&1&6&2&4\\
37&2&0&2&0&1&1&5&1&4\\
38&2&0&1&0&1&0&4&0&4\\
39&3&1&1&0&1&0&5&1&4\\
40&1&1&3&1&3&1&7&3&4\\
41&2&0&1&0&1&1&4&1&3\\
42&3&1&1&1&1&0&5&2&3\\
43&1&0&1&0&1&0&3&0&3\\
44&2&0&1&1&1&1&4&2&2\\
45&2&0&2&0&1&3&5&3&2\\
46&2&2&2&0&1&1&5&3&2\\
47&1&1&1&0&1&0&3&1&2\\
48&2&2&1&0&1&0&4&2&2\\
49&4&4&1&0&1&0&6&4&2\\
50&1&1&1&1&3&1&5&3&2\\
51&1&1&1&0&1&1&3&2&1\\
52&1&1&0&1&3&1&4&3&1\\
53&2&2&2&2&2&2&6&6&0\\
54&2&0&2&2&1&3&5&5&0\\
55&2&2&1&1&1&1&4&4&0\\
56&4&4&2&2&2&2&8&8&0\\
57&4&4&1&1&1&1&6&6&0\\
58&1&1&1&1&1&1&3&3&0\\
59&2&2&2&2&1&1&5&5&0\\
60&1&3&1&0&1&0&3&3&0\\
61&4&4&4&4&4&4&12&12&0
\end{tabular}
\caption{Table \ref{tab:real11} continued.}
\label{tab:real12}
\end{center}
\end{table}


\begin{table}
\begin{center}
\begin{tabular}{r || r r | r r | r r | r r | r }
&\multicolumn{2}{c|}{$1$}&\multicolumn{2}{c|}{$2$}&\multicolumn{2}{c|}{$3$}&\multicolumn{2}{c|}{total}&\multicolumn{1}{c}{net}\\
No.  & $+$& $-$& $+$& $-$& $+$& $-$&  $+$& $-$ &  \\\hline\hline
1&16&0&8&0&8&0&32&0&32\\
2&8&0&8&0&8&0&24&0&24\\
3&8&0&6&2&4&0&18&2&16\\
4&8&0&4&0&4&0&16&0&16\\
5&12&4&4&0&4&0&20&4&16\\
6&8&0&8&0&4&4&20&4&16\\
7&6&2&4&0&4&0&14&2&12\\
8&4&0&4&0&4&0&12&0&12\\
9&4&4&4&0&4&0&12&4&8\\
10&4&0&4&0&2&2&10&2&8\\
11&4&0&3&1&2&0&9&1&8\\
12&4&0&2&0&2&0&8&0&8\\
13&6&2&3&1&2&0&11&3&8\\
14&6&2&2&0&2&0&10&2&8\\
15&10&6&2&0&2&0&14&6&8\\
16&6&2&4&0&2&2&12&4&8\\
17&3&1&3&1&2&0&8&2&6\\
18&3&1&2&0&2&0&7&1&6\\
19&2&0&2&0&2&0&6&0&6\\
20&3&1&2&0&2&2&7&3&4\\
21&2&2&2&0&2&0&6&2&4\\
22&4&4&2&0&2&0&8&4&4\\
23&4&0&2&2&2&2&8&4&4\\
24&3&1&2&0&1&1&6&2&4\\
25&2&0&2&0&1&1&5&1&4\\
26&1&1&3&1&3&1&7&3&4\\
27&2&0&1&1&1&1&4&2&2\\
28&1&1&1&1&3&1&5&3&2\\
29&2&2&2&2&2&2&6&6&0\\
30&2&0&2&2&1&3&5&5&0\\
31&2&2&1&1&1&1&4&4&0\\
32&4&4&2&2&2&2&8&8&0\\
33&4&4&1&1&1&1&6&6&0\\
34&1&1&1&1&1&1&3&3&0\\
35&4&4&4&4&4&4&12&12&0
\end{tabular}
\caption{\cption{$E_6  \times U(1)^2 \times SO(16)$}}
\label{tab:real21}
\end{center}
\end{table}


\begin{table}
\begin{center}
\begin{tabular}{r || r r | r r | r r | r r | r }
&\multicolumn{2}{c|}{$1$}&\multicolumn{2}{c|}{$2$}&\multicolumn{2}{c|}{$3$}&\multicolumn{2}{c|}{total}&\multicolumn{1}{c}{net}\\
No.  & $+$& $-$& $+$& $-$& $+$& $-$&  $+$& $-$ &  \\\hline\hline
1&16&0&16&0&16&0&48&0&48\\
2&12&4&8&0&8&0&28&4&24\\
3&8&0&8&0&8&0&24&0&24\\
4&10&6&4&0&4&0&18&6&12\\
5&6&2&6&2&4&0&16&4&12\\
6&6&2&4&0&4&0&14&2&12\\
7&4&0&4&0&4&0&12&0&12\\
8&3&1&3&1&3&1&9&3&6\\
9&4&4&2&2&2&2&8&8&0\\
10&4&4&4&4&4&4&12&12&0\\
11&2&2&2&2&2&2&6&6&0
\end{tabular}
\caption{\cption{$E_6 \times U(1)^2  \times E_8$}}
\label{tab:real3}
\end{center}
\end{table}

\end{document}